\documentclass[nologo,url,11pt,a4paper]{ETHpaper}
\pdfoutput=1

\usepackage{amsmath, amsthm, amssymb}  
\usepackage{graphicx}      
\usepackage[numbers,sort&compress]{natbib}     
\title{Measuring Cultural Dynamics Through the Eurovision Song Contest}

\author{David Garc\'ia and Dorian Tanase}
\address{Chair of Systems Design, ETH Zurich, Weinbergstrasse 56/58\\
8092 Zurich, Switzerland\\
dgarcia@ethz.ch, dtanase@alumni.ethz.ch}

\www{\url{http://www.sg.ethz.ch}}

\begin{document}

\maketitle

\centerline{\today}

\begin{abstract}

  Measuring culture and its dynamics through surveys has important
  limitations, but the emerging field of computational social science
  allows us to overcome them by analyzing large-scale datasets.  In
  this article, we study cultural dynamics through the votes in the
  Eurovision song contest, which are decided by a crowd-based scheme
  in which viewers vote through mobile phone messages. Taking into
  account asymmetries and imperfect perception of culture, we measure
  cultural relations among European countries in terms of cultural
  affinity.  We propose the Friend-or-Foe coefficient, a metric to
  measure voting biases among participants of a Eurovision contest.
  We validate how this metric represents cultural affinity through its
  relation with known cultural distances, and through numerical
  analysis of biased Eurovision contests.  We apply this metric to
  the historical set of Eurovision contests from 1975 to 2012, finding
  new patterns of stronger modularity than using votes alone.
  Furthermore, we define a measure of polarization that, when applied
  to empirical data, shows a sharp increase within EU countries during
  2010 and 2011.  We empirically validate the relation between this
  polarization and economic indicators in the EU, showing how
  political decisions influence both the economy and the way citizens
  relate to the culture of other EU members. 

\end{abstract}

\begin{center}
\end{center}

\section{Introduction}

How do cultures evolve? How do they influence each other? These
questions are not only central to human sciences, like anthropology or
ethnology, but play a major role in politics, economics, and
international relations. Among the scientific tools to study human
cultures, agent-based modeling provides quantitative insights to
culture and opinion dynamics \cite{Castellano2009}.  The original
works of Axelrod \cite{Axelrod1997,Axelrod2006} motivate how
computational modeling can be used to understand the evolution of
human cultures. These agent-based models, while being essential
simplifications of a more complicated phenomenon, allow us to draw the
conditions for the emergence of macroscopic social behavior
\cite{Ball2007,Castellano2009}, such as polarization \cite{Flache2011}
or clustering \cite{Mas2010}, as well as to predict future social
phenomena \cite{Rybski2009,Wang2012}.  Furthermore, modeling and
simulation can open new questions that drive future research, allowing
--in an ideal case-- a deeper understanding of human societies through
multidisciplinary research \cite{Galam2004}.

As well as sociological theories need to be empirically testable,
computational models of social behavior need to be formulated over
assumptions that can be verified against empirical data. When such
behavior is objectively measurable, e.g. economic decisions
\cite{Ceyhan2011} or voting \cite{CostaFilho2003,Klimek2012}, datasets
can be produced in a way such that we can directly measure the state
of a human.  On the other hand, when a model includes subjective
elements, such as emotions or beliefs, measuring the internal states
and dynamics of a human becomes a cumbersome task. As an example,
Axelrod's model introduces the internal state of an agent as a vector
of cultural dimensions, or opinions, which change according to certain
rules \cite{Axelrod1997}. To validate this kind of dynamics, we need
to be able to measure the subjective internal state of a human, and
how it changes when interacting with others. While survey data can
shed light on opinions and culture \cite{Hofstede1980}, there is a
subconscious component of cultural behavior that cannot be encoded in
words \cite{Zizek1989}.  Nevertheless, this component can be
indirectly measured, for example through physiological responses
\cite{Kappas2011,Garcia2012}, or through online traces such as
expression biases \cite{Garcia2011}, and behavioral patterns in
computer mediated interaction \cite{Garas2012}.

Quantitative models need to be validated on the dynamics of individual
states, but are often aimed to reproduce macroscopically observable
collective behavior.  When addressing cultures or societies as a
whole, issues of data availability become critical. It can be
expensive to query large amounts of individuals, limiting the
application of subjective reports and surveys.  In addition,
approaching these questions through experimental studies suffers
additional problems. For example, experiments cannot reproduce
\emph{natural exposure} in the context of culture and popularity
\cite{McPhee1963}, limiting the representativeness of any experimental
study.  The emerging field of computational social science
\cite{Lazer2009,Giles2012} aims at overcoming these limitations,
studying human behavior through the statistical analysis of
large-scale datasets. Such datasets, when available, offer the
opportunity to validate the macroscopic behavior explained by
computational models of social interaction. Following the example of
Axelrod's model, its validation requires to measure how whole cultures
change in time, as well as the distances between different cultures.

In this article, we aim at providing a way to measure the relations
between cultures through their voting patterns in a set of song
contests, in particular looking for biases in the way they evaluate
each other. This way, we are measuring the dynamics of culture i) at a
large-scale level usually unreachable for independent research, and
ii) measuring subjective biases that are not explicitly expressed by
the studied individuals.  It is of special relevance to measure these
kind of relations in a timely manner, in order to address possible
changes in the relation between pairs of countries. The political
decisions of a country, the results of sport events, or the current
state of the economy might impact the evolution of the public opinion
of one society towards another. For the case of Europe, the policies
of the European Union regarding the debt crisis might have an impact
on the \emph{``state of the union''}, or how countries within the EU
perceive each other \cite{Ball2013}. Studying data with a time
component, we measure how these events play a role in the
manifestation of cultural relations, with the aim of providing a
macroscope that measures the state of the union of Europe at large.

\section{The Eurovision Song Contest}
In this article, we present our study of the relations between
European countries through the set of results of the Eurovision Song
Contest, an annual competition held among the country members of the
European Broadcasting Union.  Every year, each participating country
chooses a representative artist to compete by performing a song, which
is included in a live event broadcasted simultaneously in the whole
Europe. After the performance of each participant, voting countries
gather televotes and jury votes \cite{Orgaz2011}, creating a local
ranking of songs from other contestants. Afterwards, each voting
country publicly announces which other countries receive points from 1
to 8, 10, and 12, according to their local rankings. The winner of the
contest is the country with the song that accumulated the highest
amount of points.  Extensive and detailed descriptions about the
contest, its rules, and its history can be found elsewhere
\cite{Gatherer2006,Yair1996}.

While the contest rules and participating countries have changed over
the years, this contest offers a timely source of cultural evaluations
across most European countries.  Eurovision has been subject of
substantial research, up to the point of the usage of the term
\textit{``eurovisiopsephology''} \cite{Gatherer2006}, defined as the
study of the results of the votes casted in the Eurovision song
contest. Initial research focused on the possible existence of voting
clusters or alliances \cite{Yair1995,Yair1996}, generally due to
geographical locations, diaspora effects, language, and religious
similarities \cite{Spierdijk2006}. Further studies combined network
analysis with simulations of maximally random contests, revealing how
Eurovision results have high clustering \cite{Fenn2006}, which results
in voting blocks \cite{Gatherer2006,Orgaz2011}, and higher chances to
win for countries depending on their position in the voting network
\cite{Dekker2007,Saavedra2007}.

Since 2004, all the countries participating in the contest choose
their votes according to televoting, a method that uses phone calls
and mobile phone messages of viewers to decide how a country
votes. Since 2009, these televotes were combined with some expert
judges, turning Eurovision in an experimental ground to compare
popular and expert choices. Recent studies show the statistical
changes due to televoting \cite{Orgaz2011}, while older works measure
how expert judges chose their votes according to song quality rather
than cultural biases \cite{Haan2005}. Either way, the results of this
contest highlight the stable cultural relations between countries
\cite{Spierdijk2006}, where voting trades or game theoretical
decisions do not seem to play a role \cite{Ginsburgh2008}.

\subsection{Controversies and applications}
\label{sec:controversies}
Recently, Google set up a Eurovision predictor based on search
queries, leading to the correct prediction in 2009 and
2010\footnote{\url{calmyourbeans.wordpress.com/2012/05/22/no-google-eurovision-predictor-this-year/}}
. This was discontinued in 2011, after a contested prediction result
between Lena, the previous German winner who was competing again, and
the Irish participants called
Jedward\footnote{\url{wiwibloggs.com/2011/05/07/google-prediction-jedwards-lead-grows-denmark-and-estonia-climbing-update-2/10942/}}. The
outcome of this prediction failed, as both countries were defeated by
Azerbaijan by more than 100 points. In addition, seems that users were
exploiting the search engine to try to push their country higher in
the
prediction\footnote{\url{thedailyedge.thejournal.ie/google-trends-predict-eurovision-near-miss-for-jedward-130899-May2011/}},
as if searching for your representative would increase its chance to
win. This reaction to a prediction mechanism shows how social systems,
as complex adaptive systems, can change their behavior due research
results, leading to the invalidation of prediction tools or even to
self-fulfilling predictions.

Our approach does not aim to predict contest outcomes or to reveal
voting alliances, but to use Eurovision as a social macroscope for the
relations across European countries.  Initial results show how
Eurovision outcomes can predict international trade
\cite{Felbermayr2010,Kokko2012}, which motivates the measuring of the
cohesion of European countries and the EU through Eurovision
\cite{Orgaz2011}.  Popular culture and mass media criticize the
contest organization, claiming that some countries are treated as
European only in Eurovision\footnote{ ``I'm sick of being European
  just on Eurosong''
  \url{https://www.youtube.com/watch?v=IK8fVHNk0oM}}, as a limitation
for a ``Europeaniziation process''\cite{Jones2011}. In addition, the
contest rules and results are periodically claimed to be unfair,
biased\footnote{\url{http://news.bbc.co.uk/2/hi/uk_news/wales/south_east/3719157.stm}}
, or even
farcical\footnote{\url{http://news.bbc.co.uk/2/hi/entertainment/6654719.stm}},
portraying the contest as a European popularity survey rather than an
artistic competition. In this article, we precisely aim to measure
these biases as relevant quantities, focusing on the political,
social, and cultural component of the contest rather than on its
artistic one.

\subsection{Exploring Eurovision data}

We gathered the whole historical set of Eurovision results from
\texttt{Wikipedia}\footnote{For an example of a contest result page,
  see:
  \url{http://en.wikipedia.org/wiki/Eurovision_Song_Contest_2012}},
which contains a page for each edition of the contest, and from the
official website of the
contest\footnote{\url{http://www.eurovision.tv/page/history/year}}. For
each year, we count with a matrix with the values $p_{v,c}$, where
each entry corresponds to the amount of points given by a country
$c_v$ to the competing song of another country $c_c$.  As explained
before, $p_{v,c}$ is contained in the set
$\{0,1,2,3,4,5,6,7,8,10,12\}$, and chosen according to the ranking of
televotes and jury votes.  Our dataset comprises the whole set of
results of Eurovision editions from 1957 to 2012, including 11775
voting relations between 51 country members of the European
Broadcasting Union.

\begin{figure}[th]
\centerline{\includegraphics[width=0.52\textwidth]{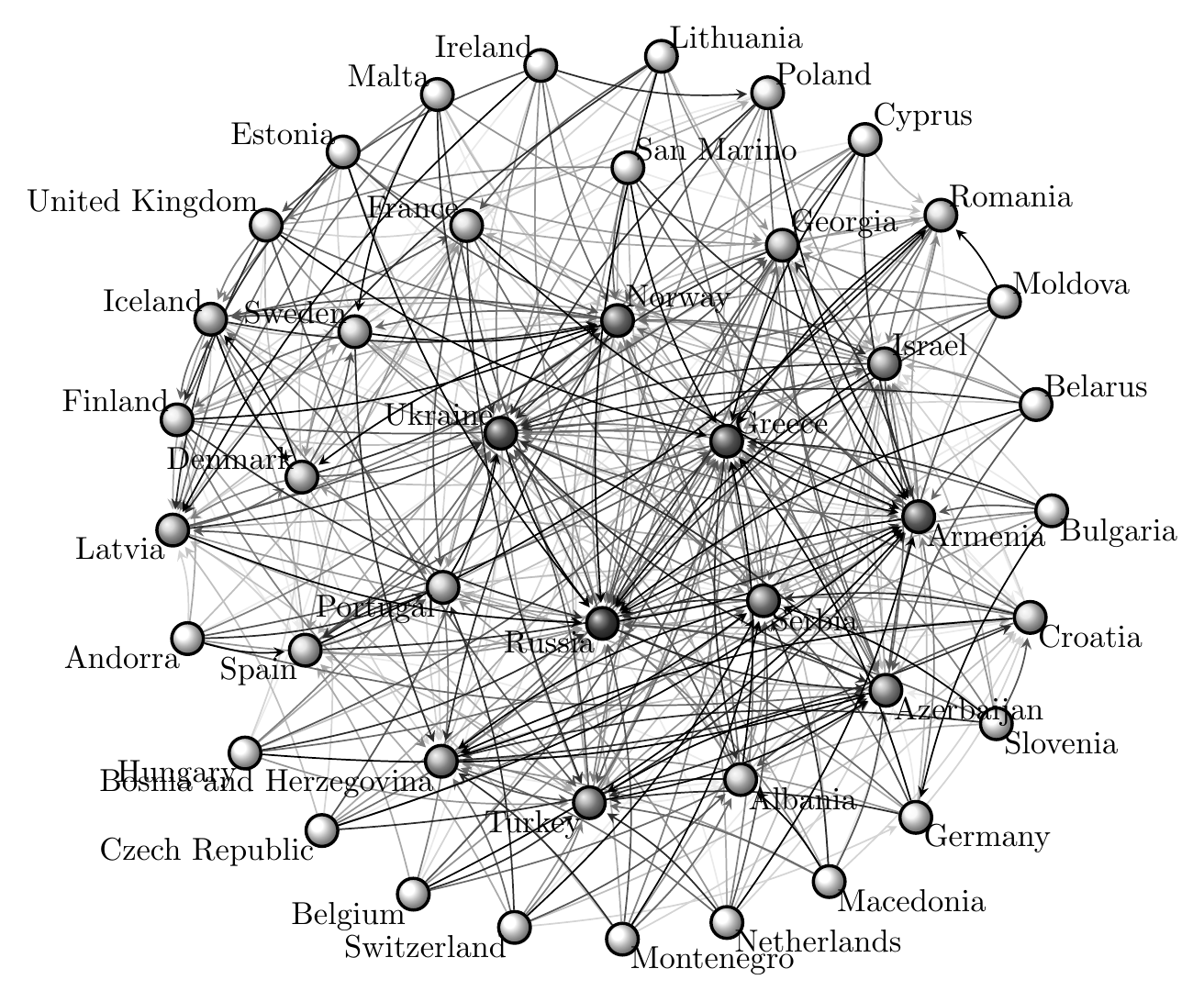} 
} \vspace*{8pt}\caption{Network of votes for the 2008 edition of
  Eurovision. Nodes are participating countries, with color darkness
  proportional to the final score of the country.  Directed edges
  represent the votes given by one country to another, with darker
  color according to the amount of points given by the
  vote.\label{fig:voteNets}}
\end{figure}

The straightforward approach to understand this data is to look into
the voting network formed every year \cite{Fenn2006}, where nodes are
participating countries. A directed edge $c_v \rightarrow c_c$
connects two nodes if $c_v$ assigned that year a nonzero amount of
points to $c_c$. Edge weights are assigned to be the amount of points
given by the vote, $p_{v,c}$.  Figure \ref{fig:voteNets} shows this
network for the edition of 2008, with edge darkness according to
weight, and node darkness proportional to the final score $s_c =
\sum_{c_v} p_{v,c}$ of each country $c_c$ in the contest. The
topological properties of these networks have been widely explored,
finding symmetrical relations, triadic clustering, and highly
connected blocks that map to geographically close, and culturally
related countries \cite{Yair1996,Saavedra2007,Gatherer2006}.

\begin{figure}[th]
\centerline{
\includegraphics[width=0.6\textwidth]{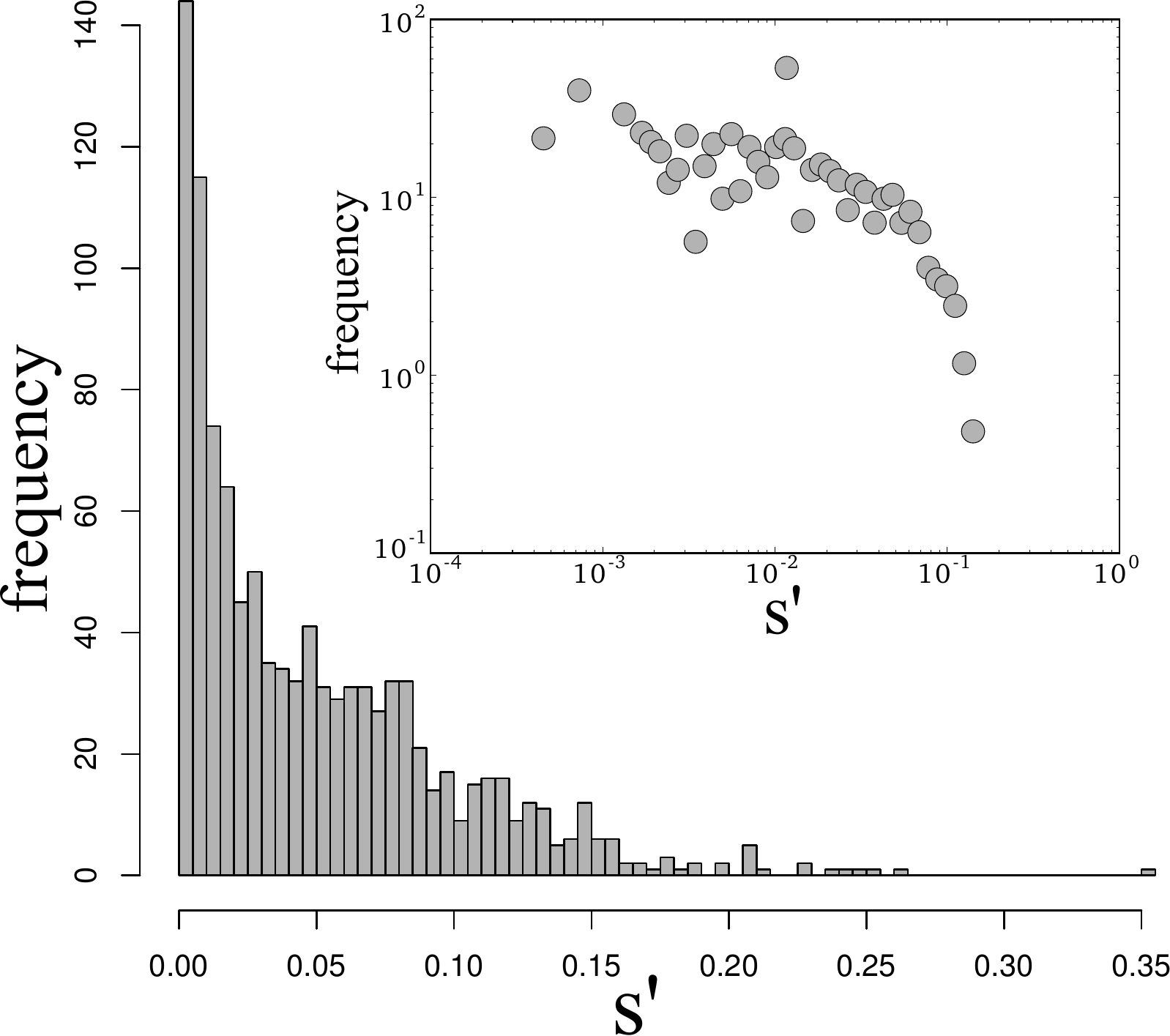}}
\vspace*{8pt}
\caption{Relative score $s'$ distribution for contest results from
  1975 to 2012. Inset: log-log version of the distribution. We use
  this empirical distribution as input for our
  simulations. \label{fig:scoreDists}}
\end{figure}

Visual inspection of this network, as shown in Figure
\ref{fig:voteNets}, reveals a significant heterogeneity in node
darkness. This corresponds to the large deviation of final scores
usually present in this contest. Initial editions of the contest had
multiple draws, so the voting scheme was changed to the current one in
1975, in order to encourage the selection of a single winner. The
resulting heterogeneity is relevant to test the existence of
winner-takes-all effects as in cultural markets \cite{Rosen1981}, and
product reviews \cite{Leskovec2007}. To do so, we calculated the
relative score $s'_{c}= \frac{s_c}{T}$, where $T=\sum_{c_c}s_c$ is the
total amount of points given in an edition of the contest, which
depends on the amount of countries participating in a given year. This
way we can aggregate all participant scores since 1975, as shown in
the histogram of Figure \ref{fig:scoreDists}. Similarly to the
cultural markets mentioned above, the distribution of $s'$ shows a
large variance, and positive skewness. On the other hand, the log-log
histogram shown in the inset of Figure \ref{fig:scoreDists} allows us
to notice that there are no scaling relations, probably due to the
finite size of the contest. We can say that the contest has a large
variance of final scores, yet these do not allow arbitrarily large
values, as opposed to previous experience in popularity analysis. We
will use these final scores to compare individual votes with final
results, as explained below.

\section{Measuring cultural relations through Eurovision}
\subsection{Perception of culture}
\label{sec:theory}
Most agent-based models of culture dynamics include agent interactions
based on their internal states, usually depending on the distance
between the values of their cultural features. While valid to
reproduce the emergence of opinion groups and cultures
\cite{Michard2005}, there is still ample room to validate and
empirically test the existence of this kind of dynamics. The existence
of cultural dimensions was first introduced by Hofstede
\cite{Hofstede1980}, in a study of surveys across different countries.
These dimensions were detected by means of dimensionality reduction on
survey responses, and have been applied in numerous studies about
culture \cite{Fernandez1997}, including a study on Eurovision
\cite{Ginsburgh2008}. On the other hand, measuring culture through
surveys has clear limitations \cite{Ailon2008}, in particular in the
interpretation of the meaning of the results of dimensionality
reduction.

Apart from dimensional structures, a key component in models of
culture dynamics is the set of rules that determine which agents
interact and how. While these rules can represent spontaneous events
of influence between cultures, in other scenarios work as a mechanism
in which agents perceive the state of others.  In a realistic setup,
the perception of cultural differences might be constrained by
imperfect communication, and path dependencies like historical events
or stereotypes. Such phenomena can shape the way culture is perceived
across a society, leading to new structures to take into account in
future models.  It is important to highlight that these structures and
dynamics depend on the societal level at which culture is defined,
which can include countries \cite{Hofstede1980}, ethnic groups
\cite{Atkinson2011}, or firm clusters \cite{Groeber2009}.

\begin{figure}[th]
\centerline{\includegraphics[width=0.6\textwidth]{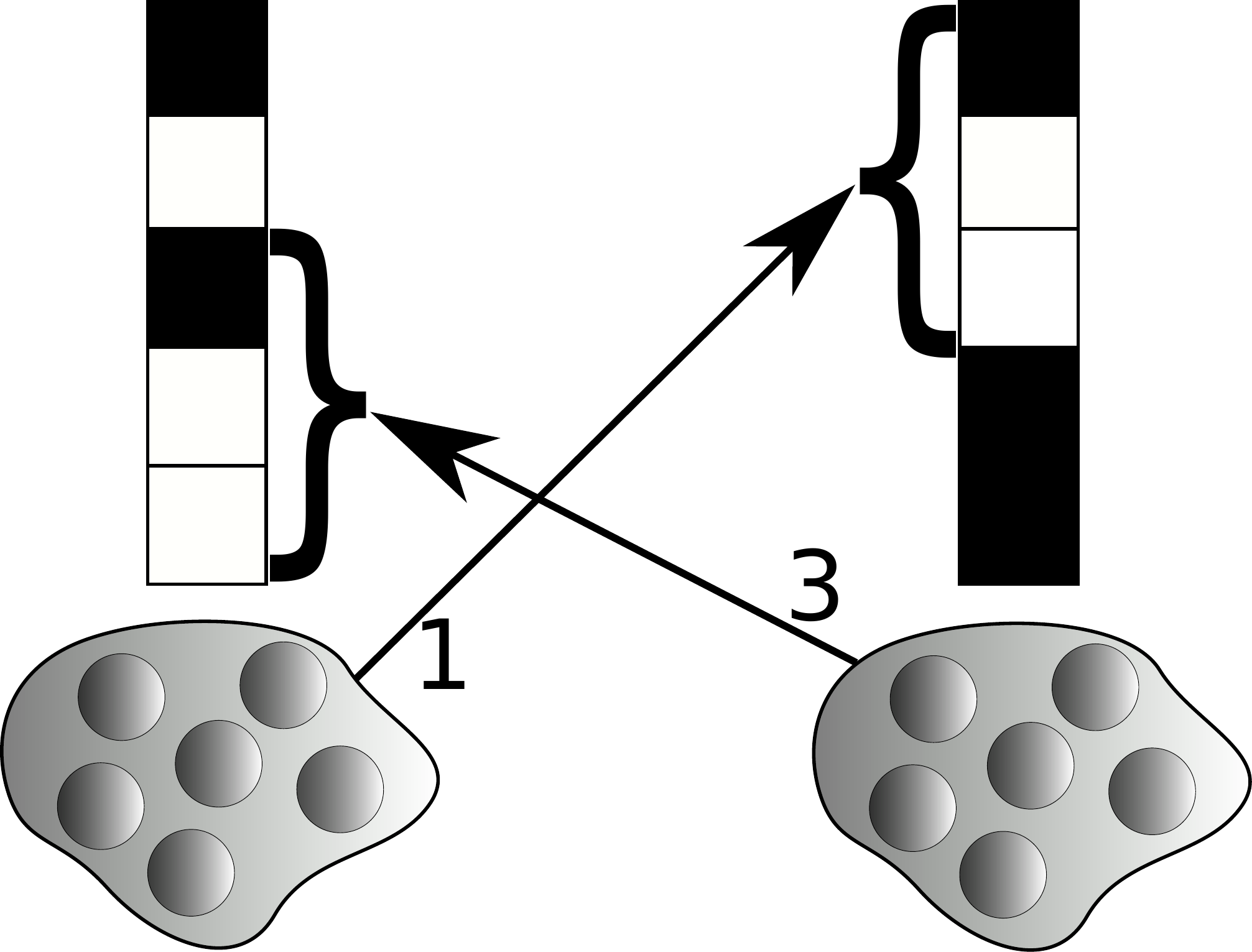}}
\vspace*{8pt}
\caption{Schema with a possible scenario of feature-based cultures
  under imperfect perception. Each culture is composed of a set of
  agents with the same values of the cultural vector, shown over
  them. Their asymmetric relation is a result of partial perception of
  the cultural features of the other. \label{fig:schema}}
\end{figure}

With the minimal assumption that humans can only perceive a set of
dimensions from another culture, the perceived distance between
cultures could have asymmetric properties. In the schema of Figure
\ref{fig:schema}, we sketch two cultures with binary feature vectors
of five dimensions. If the left one can only perceive the three first
features of the other, its perceived Hamming distance would be 1, as
they just differ in the third feature. At the same time, if the right
one can only perceive the last three features of the left one, the
perceived distance would be 3, leading to asymmetric perception of
cultural differences.

Another possibility is that each society might have a reference point,
i.e. an expected or ``acceptable'' maximum cultural distance towards
another. This would lead to the existence of negative cultural
relations, which would be a plausible explanation for multiple
international conflicts present in History. This possibility is
commonly ignored when taking into account cultural distances in
discrete spaces, and might very well be a property of realistic
cultural dynamics.

Given possible asymmetries and signed values, we will define
\emph{cultural affinity} of a society towards another as \emph{``the
  differences perceived by the members of a society in relation to the
  culture of another society, evaluated as a comparison with a
  reference point''}. Cultural affinity takes maximum values for very
close cultures, and negative values towards very different cultures.

In this article, we analyze cultures at the country level, aggregating
individuals according to the country in which they live.  By analyzing
Eurovison results, we want to explore the asymmetry in the perceived
differences between cultures, and the existence of negative and
positive cultural evaluations among the cultures of European
countries. This datasource implies a salience of musical tastes as a
feature of culture. Nevertheless, we will focus on the patterns of
cultural affinity to explore the relation between Eurovision votes and
other factors, including Hofstede's cultural dimensions and economic
relations.

\subsection{The Friend-or-Foe coefficient}
To measure cultural affinity, we need to define a way to estimate it
from the raw Eurovision scores of our dataset. For this, we define the
Friend-or-Foe (FoF) coefficient of country $c_v$ towards country $c_c$, as
estimated from a particular edition of the contest:
\begin{equation}
Fof(c_v, c_c) = \frac{p_{v,c}}{12} - \frac{s_c - p_{v,c}}{12(N-2)}
\label{eq:fof}
\end{equation}
where $p_{v,c}$ are the points assigned to $c_c$ by $c_v$, $s_c$ is
the final score of $c_c$, and $N$ is the total amount of countries
voting in the studied edition of Eurovision. The first term of the
right hand side of Equation \ref{eq:fof} represents a normalized value
of the score given by $c_v$ to $c_c$, ranging from $0$ for
no points given, to $1$ when 12 points were assigned to $c_c$. The
second term corrects for the final score of $c_c$ in the whole
contest, calculating the total amount of points given by \emph{other}
countries different than $c_v$. The maximum value of this score is
$12(N-2)$, as one country cannot vote itself and we have already
subtracted $c_v$ from the calculation.

We designed the Friend-or-Foe coefficient to measure the overvoting or
undervoting bias from a country to another, correcting for ``song
quality'' as estimated by the final contest result
\cite{Ginsburgh2008}. This way, we aim at removing the effects of the
artistic component of the contest, highlighting the political or
cultural biases that are commonly claimed to exist in Eurovision. If a
country $c_v$ assigns 12 points to $c_c$, while all the others assign
0, then $FoF(c_v,c_c)=1$, which would be the maximum value of an
overvoting bias. If $c_v$ assigns 0 points to $c_c$ but all the other
countries assign 12, then $FoF(c_v,c_c)=-1$, representing the
maximally negative Friend-or-Foe coefficient given the contest rules.

After this definition, we need to assess if the FoF is a valid measure
to estimate the real cultural affinity of one country towards another,
as described above.  In the following, we test the relation between
the FoF and previously know cultural distances between European
cultures. Furthermore, we explore some examples of country pairs with
known cultural similarities, as well as countries with explicit
conflicts.

\subsection{Relation between culture and FoF}
\label{sec:caseStudy}

For any pair of countries $c_1$ and $c_2$ we can calculate the FoF
coefficients between them in each contest in which they participated
together. The values of $FoF(c_1,c_2)$ and $FoF(c_2,c_1)$ might depend
on effects that influence Eurovision votes, including cultural
affinity. In this section, we show the relation between culture and
the FoF, using two metrics on independent datasets:

\begin{itemize}
\item \textbf{Mean FoF.} We aggregate the FoF coefficients for each
  pair of countries through $\overline{FoF}(c_1,c_2) =
  \frac{1}{M_{c_1,c_2}} \sum_{t} FoF_t(c_1,c_2) $ where
  $FoF_t(c_1,c_2)$ is the FoF between $c_1$ and $c_2$ on year $t$ and
  $M_{c_1,c_2}$ is the amount of times both countries participated
  together since 1975.

\item \textbf{Cultural distance.} We take as a ground truth the
  quantization of cultural distances provided by Hofstede
  \cite{Hofstede1980}\footnote{These values can be browsed at
    \url{http://geert-hofstede.com/dimensions.html}}, including 17
  countries that have participated in Eurovision\footnote{Austria,
    Belgium, Denmark, Finland, France, Germany, Greece, Ireland,
    Israel, Netherlands, Norway, Portugal, Spain, Sweden, Switzerland,
    Turkey, United Kingdom}. This way, for each country $c$, we have
  measures of four different cultural dimensions: \emph{Power
    Distance} $p_c$, \emph{Individualism} $i_c$, \emph{Masculinity}
  $m_c$, and \emph{Uncertainty Avoidance} $u_c$.  We calculate the
  cultural distance between the countries $c_1$ and $c_2$ as
\begin{equation}
d(c_1, c_2) = \frac{1}{100} ( |p_{c_1} - p_{c_2}| +
  |i_{c_1} - i_{c_2}| + |m_{c_1} - m_{c_2}| + |u_{c_1} - u_{c_2}|)
\label{eq-distance}
\end{equation}
which corresponds to the Manhattan distance between both cultures,
rescaling each dimension.
\end{itemize}

\begin{figure}[th]
\centerline{\includegraphics[width=0.6\textwidth]{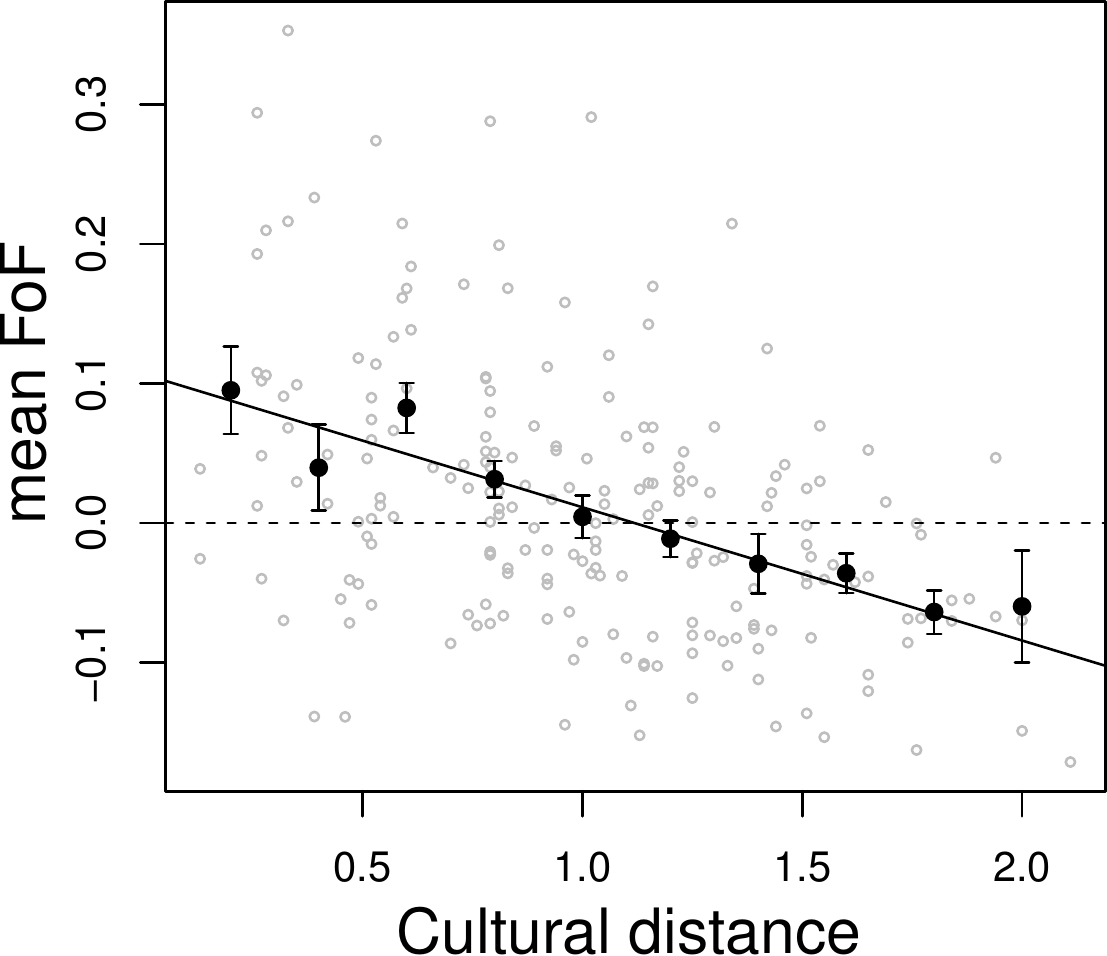}}
\vspace*{8pt}
\caption{Scatter plot of the mean Friend-or-Foe coefficient
  (1975-2012) versus Hofstede's cultural distance between countries
  that participated together in Eurovision more than 25 times since
  1975.  Barplot shows mean and standard error in 10 bins, and the
  solid line shows the linear regression result ($R^2= 0.1946,
  p<10^{-10}$).\label{fig:Culture-FoF}}
\end{figure}

These two metrics are constrained by the availability of data, as we
need enough values of the FoF to calculate their mean. Taking pairs
that co-participated in Eurovision more than 25 times leaves us with a
total of 206 country pairs, which account for more than 75\% of all
the possible pairs of countries included here. Figure
\ref{fig:Culture-FoF} shows a scatter plot of
$\overline{FoF}(c_1,c_2)$ versus $d(c_1,c_2)$, with superimposed mean
values of $\overline{FoF}(c_1,c_2)$ in 10 bins of $d(c_1,c_2)$. There
is a pattern of declining $\overline{FoF}(c_1,c_2)$ for country pairs
with higher cultural distances. The Pearson's correlation coefficient
between the mean FoF and the cultural distance is $-0.441$, in a 95\%
confidence interval of $[-0.545,-0.324]$, and with $p<10^{-10}$.

Additionally, we calculated the linear regression of
$\overline{FoF}(c_1,c_2)$ as a function of $d(c_1,c_2)$, finding that
the weight of cultural distance is significant and estimated as
$-0.09558$. The result of this linear regression is shown as a solid
line in Figure \ref{fig:Culture-FoF}, which crosses $0$ at a point
close to distance 1. This suggests the existence of a reference point,
beyond which we can expect two countries to undervote each other,
i.e. have negative FoF if their cultures are at a distance above 1.

To gain deeper knowledge on this relation between cultures and the
FoF, we show four examples of country pairs and their FoF, shown in
Figure \ref{fig:FoFs}. These examples illustrate three properties of
the FoF:

\begin{figure}[th]
  \centerline{\includegraphics[width=0.48\textwidth]{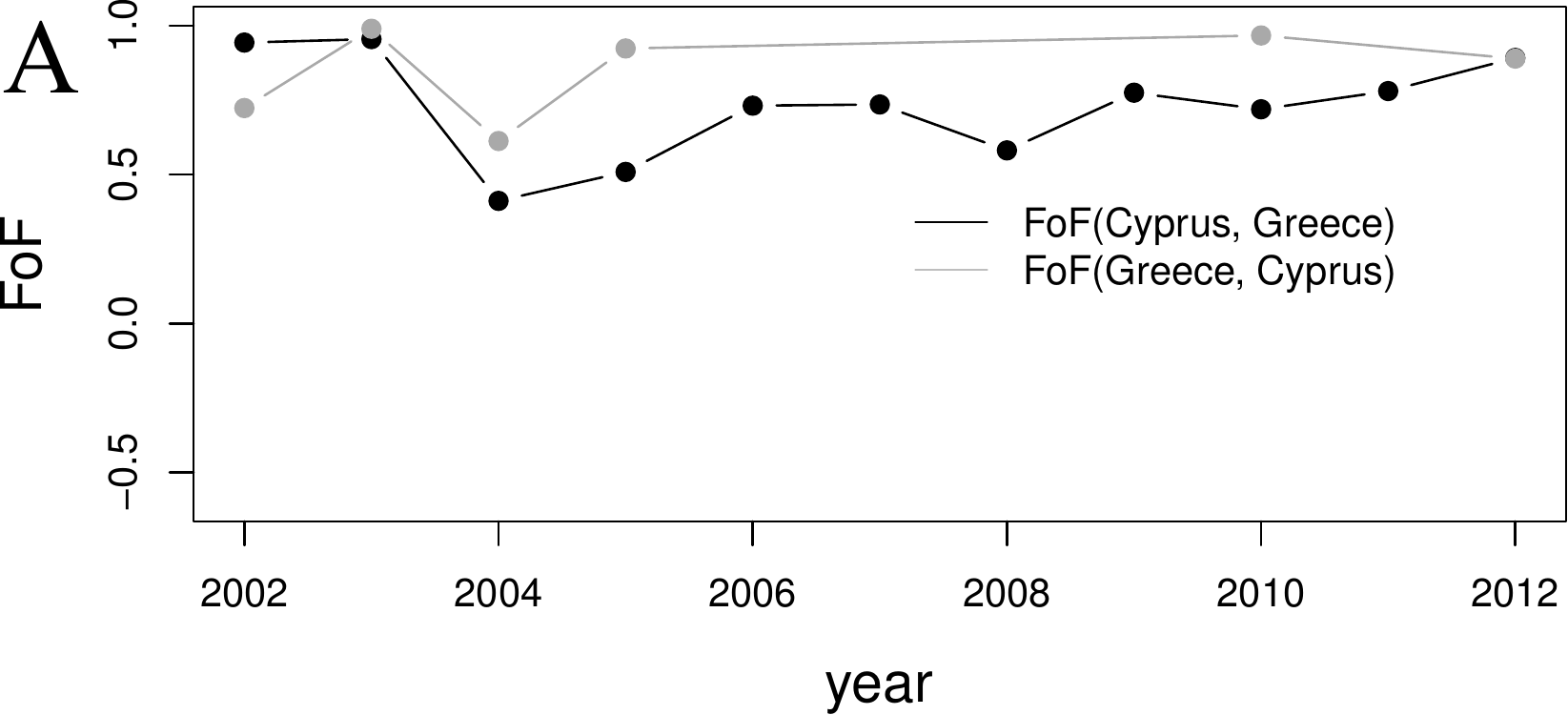}
  \hfill \includegraphics[width=0.48\textwidth]{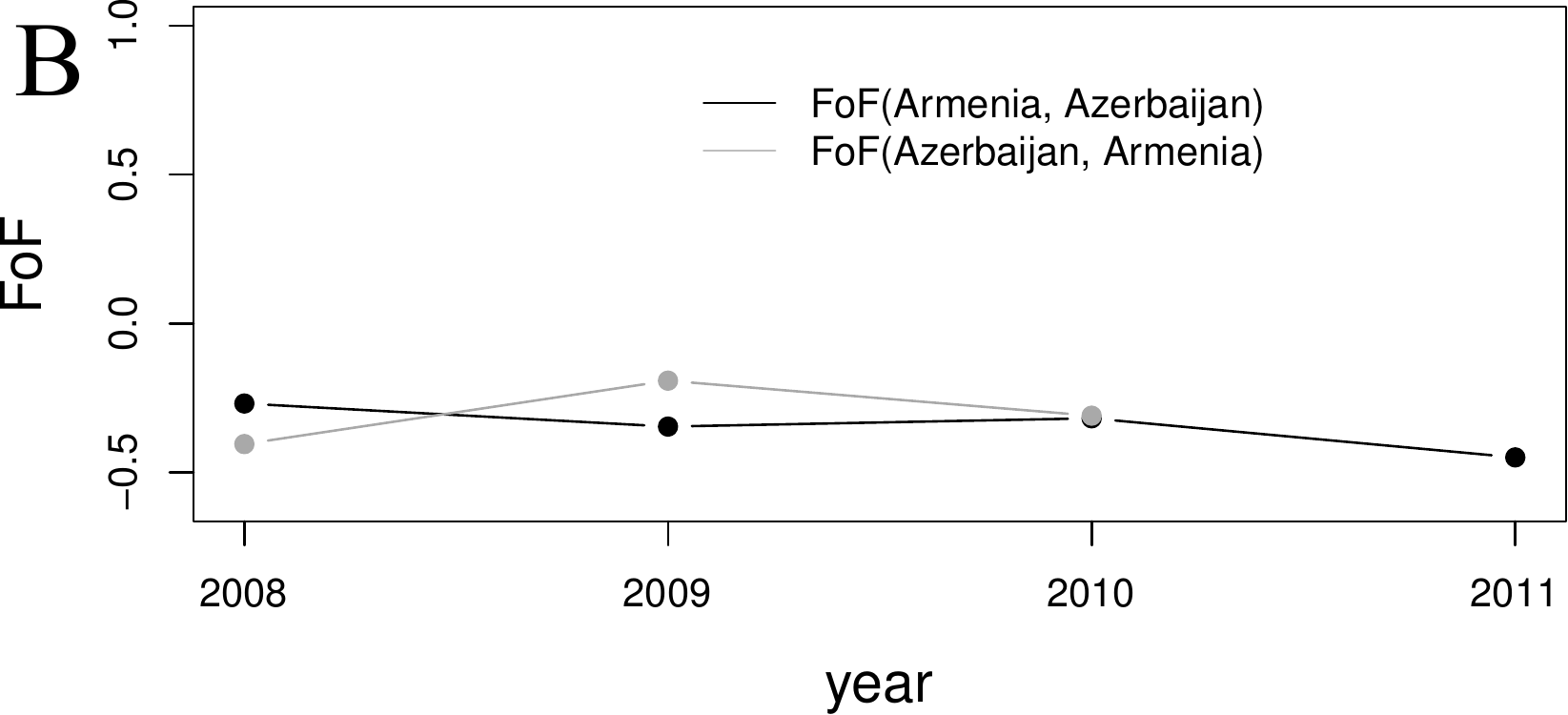}}
\centerline{ \includegraphics[width=0.48\textwidth]{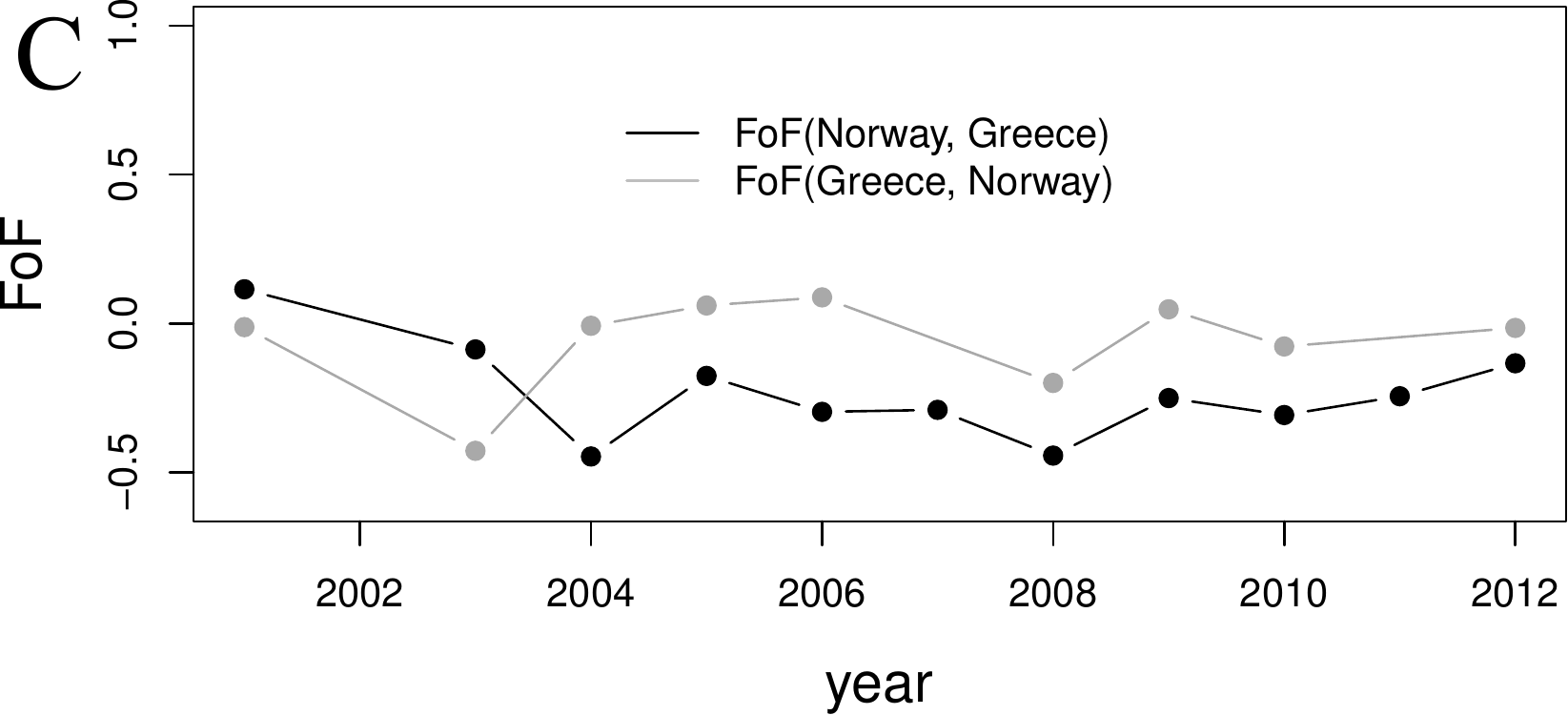}
   \hfill \includegraphics[width=0.48\textwidth]{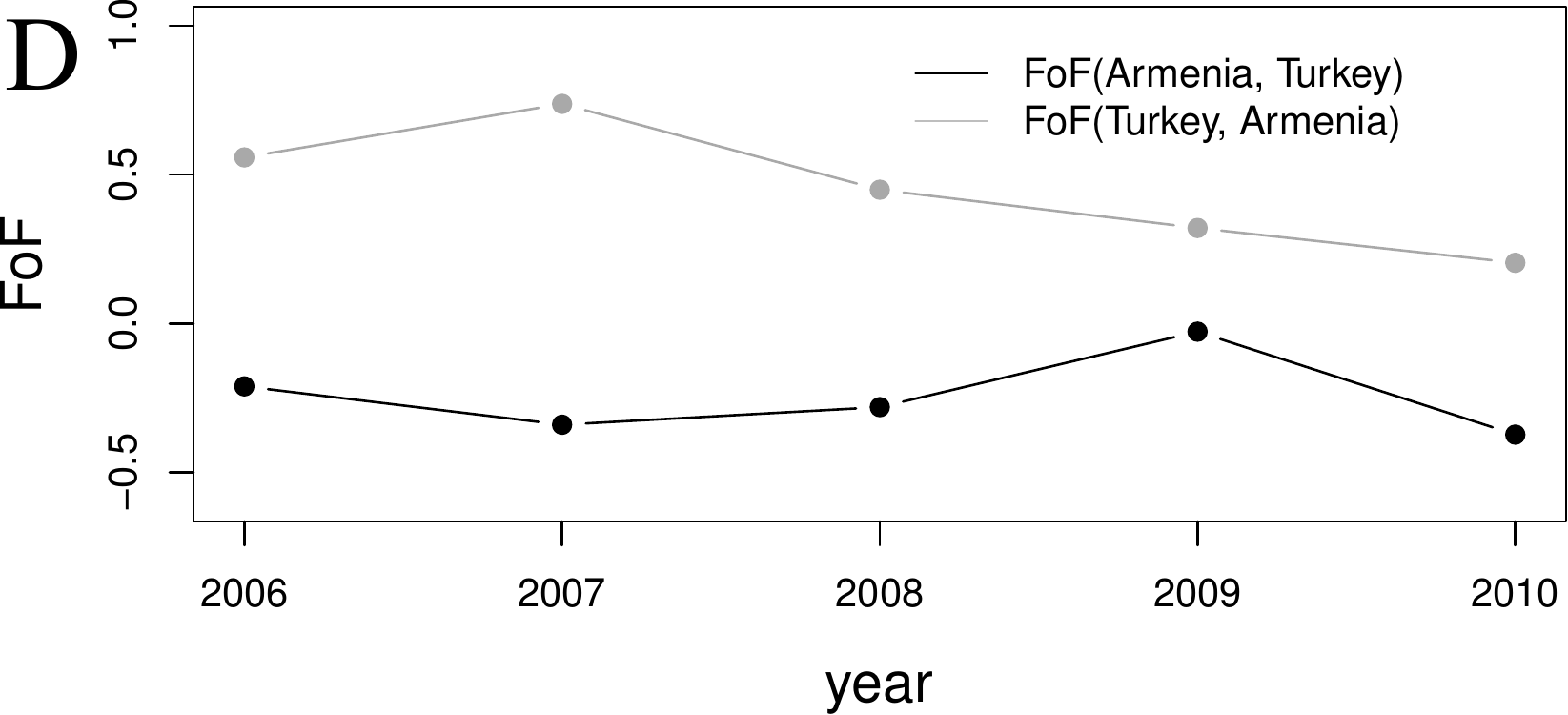}}
\vspace*{8pt}
\caption{Examples of yearly Friend or Foe coefficients, for the pairs
  A: (Cyprus, Greece), B: (Armenia, Azerbaijan), C: (Norway, Greece),
  and D: (Armenia, Turkey). \label{fig:FoFs}}
\end{figure}

\begin{itemize}
\item \textbf{Cultural proximity.} The standard example for the
  expression of cultural similarity in Eurovision is Cyprus and Greece
  \cite{Fenn2006,Gatherer2006,Orgaz2011}.  Figure \ref{fig:FoFs}A
  shows the FoF between these two countries from 2002 until 2012. Both
  values are positive in each edition of the contest, never dropping
  below 0.3. This example is in line with our above analysis of the
  relation between FoF and cultural distance, were Cyprus and Greece
  is an extreme point with very low distance.  

\item \textbf{Asymmetric effects.}  The FoF pair of Turkey and
  Armenia, shown in Figure \ref{fig:FoFs}D is a clear example of an
  asymmetric relation between countries. $FoF(Turkey, Armenia)$ keeps
  a significantly positive value, hypothetically due to Armenian
  diaspora living in Turkey. This same effect of 'patriotic voting'
  was suggested for Turkish migrants across Europe
  \cite{Spierdijk2006}, and our FoF coefficient reflects it in this
  case. On the other hand, $FoF(Armenia, Turkey)$ is significantly low
  and mostly below 0. This negative relation is a possible expression
  of negative relations due to historical conflicts between both
  countries. We analyze these asymmetries at a global level in our
  modularity analysis of Section \ref{sec:modularity}.

\item \textbf{Negative relations.} We want to explore the possibility
  of negative relations between pairs of countries. Couples of
  countries with explicit territorial conflicts show this symmetric
  negative FoFs, as shown in Figure \ref{fig:FoFs}B.  Armenia and
  Azerbaijan are still officially in war since the Nagorno-Karabakh
  conflict\footnote{\url{http://en.wikipedia.org/wiki/Nagorno-Karabakh_War}}. This
  negativity is evident in their FoF coefficients, as these countries
  consistently avoid voting each other.  The example of Greece and
  Norway in Figure \ref{fig:FoFs}C shows that negative FoF can
  indicate negative cultural affinity as well. The cultural distance
  between Norway and Greece is 1.74, placing this pair as one of the
  most distant cultures, beyond the reference point of 1, and thus
  having negative FoF coefficients.
\end{itemize}

\section{Analysis and simulation of biased contests}

\subsection{Properties of the Friend-or-Foe coefficient}

The FoF is an indirect measure of the underlying cultural affinity
between countries, and the contest rules or the artistic component of
the performances can influence this metric. In this section, we
provide a detailed analysis of the FoF coefficient based on
simulations, comparing the influence of contest rules and cultural
affinities in the Friend-or-Foe coefficient.

By definition, the mean of the FoF values directed to a particular
country is 0:
\begin{equation} 
\sum_{c_v} FoF(c_v, c_c) = \sum_{c_v} \frac{p_{v,c}}{12} - \frac{s_c - p_{v,c}}{12(N-2)}
=-(N-1)\frac{s_c}{12(N-2)} + \sum_{c_v} \frac{(N-1) p_{v,c} }{12(N-2)}
\end{equation}, $\sum_{c_v} p_{v,c} = s_c$ and there are $N-1$ countries that can
vote $c_c$, leading to a total sum of 0. According to this property,
the expected value of the FoF between random pair of countries is 0. Beyond
this zero mean, we can expect the distribution of FoF for a contest to be
far from uniform, and not all $s_c$ and $p_{v,c}$ are equally likely
in the empirical data.

\begin{figure}[th]
\centerline{
\includegraphics[width=0.58\textwidth]{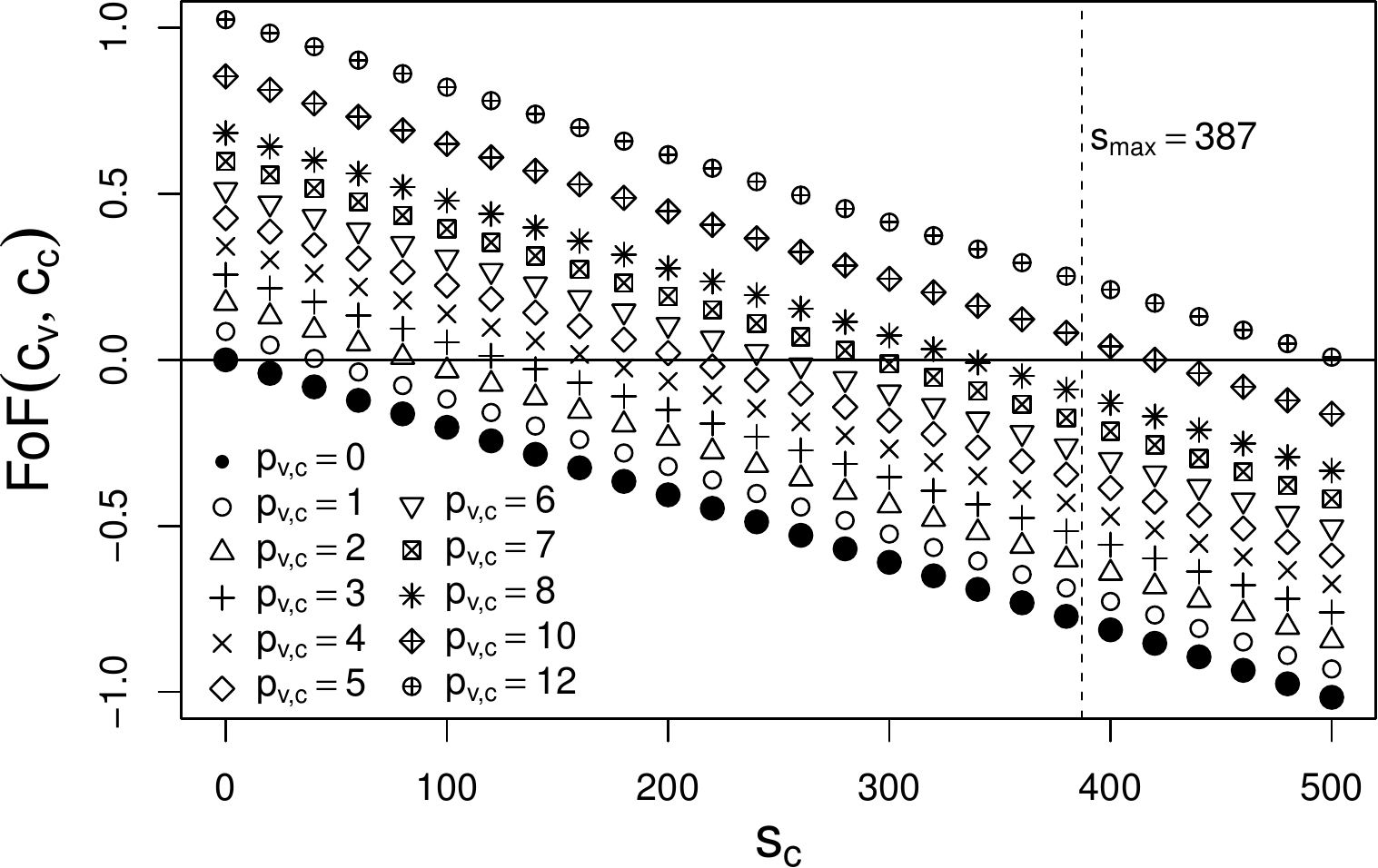}
\hfill
\includegraphics[width=0.4\textwidth]{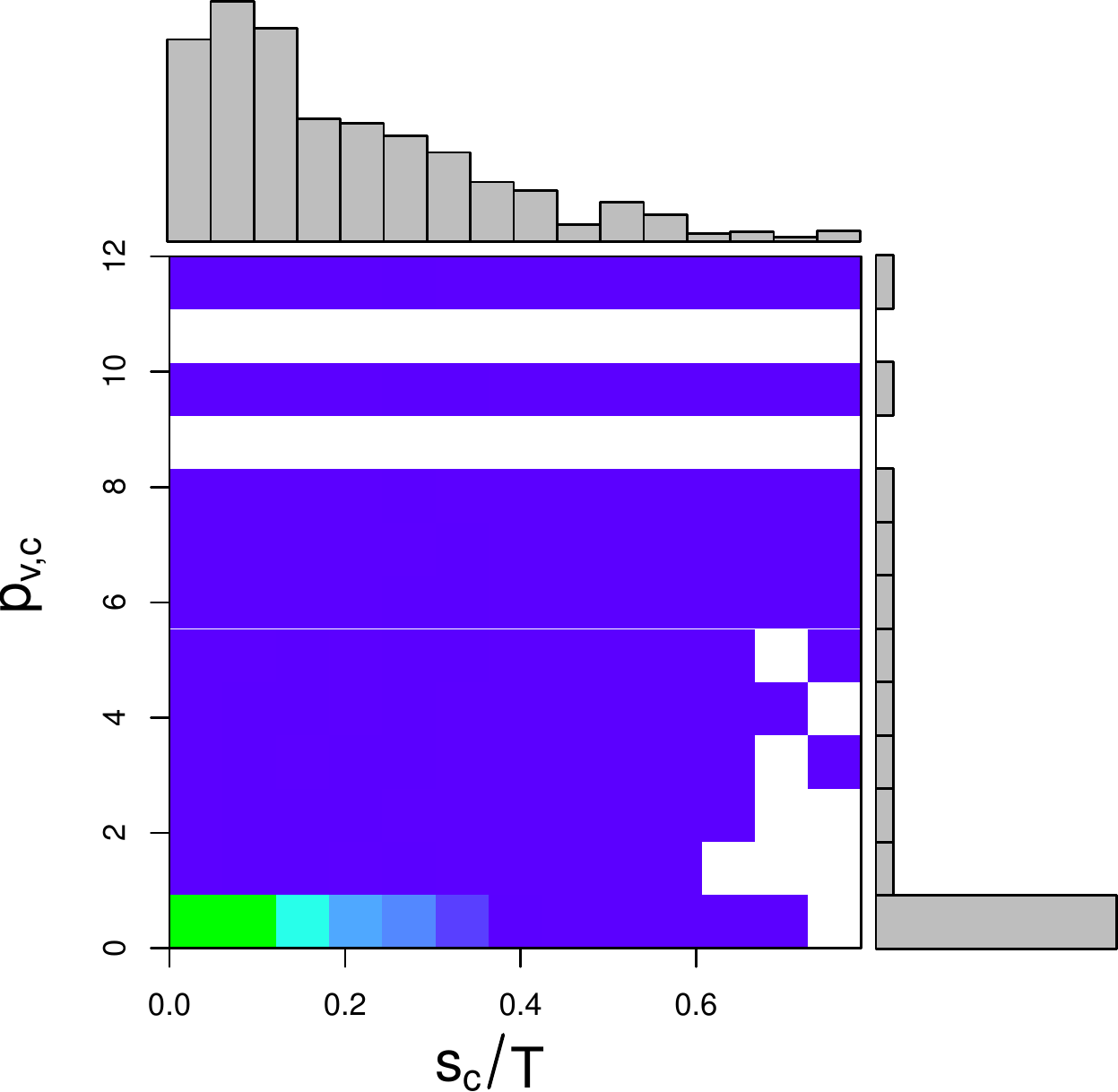}}
\vspace*{8pt}
\caption{Left: $FoF(c_v,c_c)$ values corresponding to the different
  possible combinations of $p_{v,c}$ and $s_c$, in a contest with
  $N=43$. Right: histogram of empirical values of pairs $p_{v,c}$,
  $s_c/T$ for contests in the period 1997-2012, ranging from
  blue to green proportionally to density. Barplots show
  the histograms of $s_c/T$ (top), and $p_{v,c}$ (right).
  \label{fig:FofAnalysis}}
\end{figure}

The FoF increases linearly with the score given from one country to
another, and decreases with the final score of the voted country. The
left panel of Figure \ref{fig:FofAnalysis} shows the value of the FoF
for different combinations of $p_{v,c}$ and $s_c$, in a contest with
43 voting countries. The possible FoF values allows a range from 1
when a country gives 12 points to another with a final score of 12, to
-1 when a country gives 0 points to another that got 12 points from
each other country. Both cases are possible, but the latter seems much
less likely under the skewed distribution of final results shown in
Figure \ref{fig:scoreDists}. In particular, the maximum amount of
points ever achieved by a participant in Eurovision is $s_{max}=387$,
added as a vertical line in Figure \ref{fig:FofAnalysis}. Without
additional knowledge of the contest, this would suggest that the FoF
has a tendency towards positive values, as there would be more
possible combinations of $s_c$ and $p_{v,c}$ giving a FoF above 0.

The skewness of the final scores is not the only factor that can
shapes the FoF. Due to the contest rules, voting countries have a
fixed voting scheme, which assigns the fixed values of
$[1,2,3,4,5,6,7,8,10,12]$ to 10 other participants, and $0$ to all the
rest. This implies that a value of $p_{v,c}=0$ is much more likely
than any other, i.e. the black dots of Figure \ref{fig:FofAnalysis}
would appear more often in a contest. Given $s_c>0$, a $p_{v,c}=0$
would imply a negative FoF, suggesting a negative tendency in the FoF
values as opposed with the positive one explained above.

\begin{figure}[th]
\centerline{\includegraphics[width=0.95\textwidth]{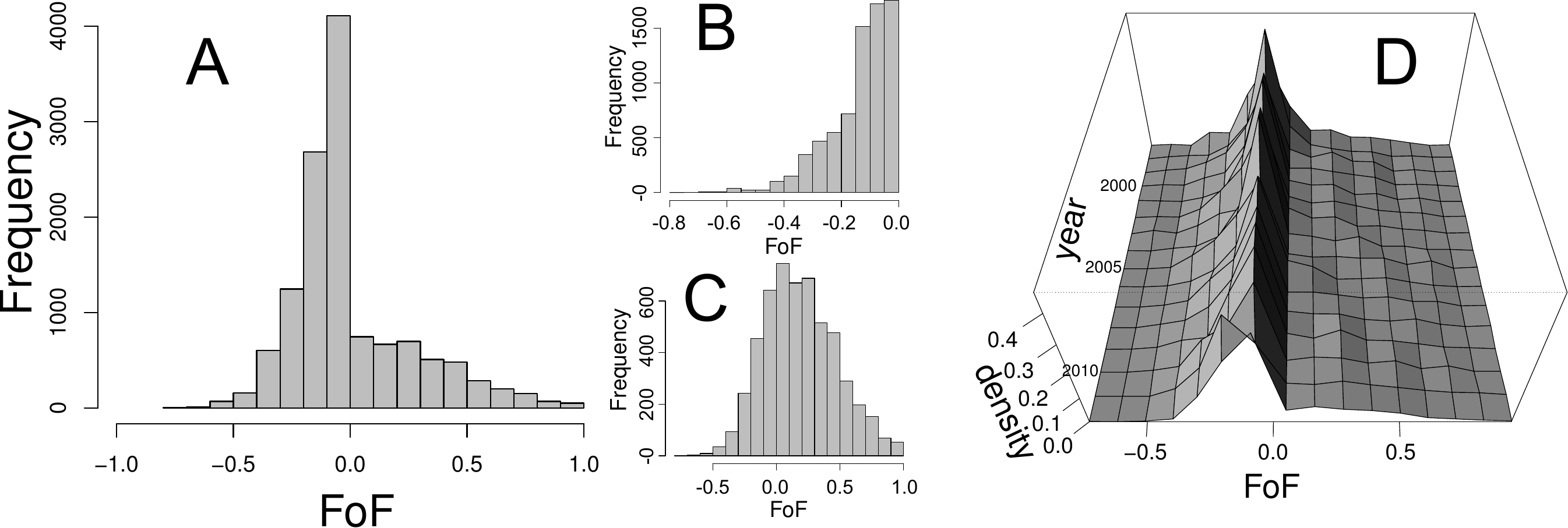}}
\vspace*{8pt}
\caption{A) Histogram of FoF values for all contests between 1997 and
  2012.  B) Histogram of FoF when $p_{v,c}=0$, and C) when
  $p_{v,c}>0$. D) Distribution of the FoF values for each yearly
  edition of Eurovision between 1997 and 2012.
\label{fig:FofEmpirics}}
\end{figure}

The combination of these two statistical effects can be seen in the
right panel of Figure \ref{fig:FofAnalysis}, in a histogram of the
combination of possible votes $p_{v,c}$ and rescaled final scores
$s_c/T$, for all contests from 1997 until 2012. We focus on this
period since it was in 1997 when televoting introduced, adding
additional social value to the Eurovision data.  First, the rescaled
final scores are very far from being uniformly distributed, indicating
the positive FoF tendency explained above. Second, most of the
$p_{v,c}$ values are $0$, suggesting the negative FoF tendency created
by the rules of the contest. The historical distribution of FoF shows
these effects, as shown in Figure \ref{fig:FofEmpirics}A. Negative FoF
values are more likely, but are also smaller in magnitude than
positive ones, giving the a mean of zero. The effect of the voting
scheme can be seen in Figure \ref{fig:FofEmpirics}B and C, where we
show the histogram of FoF when $p_{v,c}=0$ in the former, and when
$p_{v,c}>0$ in the latter. A score of 0 ensures a maximum FoF of 0,
while a score above 0 allows both positive and negative FoF. On the
other hand, the FoF when $p_{v,c}>0$ have a positive mean, as high
final scores happen only for very few participants.

In addition, the shape of the distribution of FoF does not
significantly change over the year. Figure \ref{fig:FofEmpirics}D
shows the distributions of FoF from 1997 to 2012, revealing that
different contest sizes do not create additional biases.

\subsection{Modeling Eurovision contests}
\label{sec:nullmodels}
The above empirical analysis shows the existence of biases in the FoF,
but it still can contain relevant information about the cultural
affinity across countries. In the following, we show a numerical
comparison of simulated Eurovision contests, increasingly including
the contest rules, heterogeneity in song quality, and cultural
affinities.  We start defining a null model, in which countries can
freely vote other countries with the only restriction of assigning a
fixed amount of points:

\fbox{
\parbox{0.85\textwidth}
 {\textbf{Null model}\\
  For each voting country $c_v$:
  \begin{enumerate}
  \item For each competing country $c_c$:
    \begin{itemize}
    \item Sample $fit_v[c_c]$ from uniform distribution between 0 and 1
    \end{itemize}
  \item For each competing country $c_c$:
    \begin{itemize}
    \item Assign $p_{v,c} = 58 * \frac{fit_v[c_c]}{\sum_c  fit_v[c]}$
    \end{itemize}
  \end{enumerate}
}}

In the null model, countries choose the votes they assign to other
countries at random, under the unique constraint that the sum of
scores they can assign is a fixed value. We used this null model as a
reference point to quantify the effect that the voting scheme has on
the distribution of FoF. To do so, we use the minimal model of random
votes under the voting scheme of Eurovision, introduced in
\cite{Gatherer2006}:

\fbox{
\parbox{0.85\textwidth}
 {\textbf{Model 1}\\
  scores = $[12,10,8,7,6,5,4,3,2,1]$\\
  For each voting country $c_v$:
  \begin{enumerate}
  \item For each competing country $c_c$:
    \begin{itemize}
    \item Sample $fit_v[c_c]$ from uniform distribution between 0 and 1
    \end{itemize}
  \item For each competing country $c_c$:
    \begin{itemize}
    \item If $rank(fit_v[c_c]) \leq 10$:  assign $p_{v,c} = scores[rank(fit_v[c_c])]$
    \item Else: assign $p_{v,c} = 0$
    \end{itemize}
  \end{enumerate}
}}
 
Previous works compared the network properties of the votes in this
model with empirical data \cite{Fenn2006}. In our study, we use this
model to assess the effect of the voting scheme in the distribution of
FoF. For each contest between 1997 and 2012, we run 100 simulations of
the null model and model 1, using the same amount of voting and
competing countries. We computed the FoF values for each of the
simulations, producing a simulated dataset of FoF values 100 times
larger than the empirical FoF data.

\begin{figure}[th]
\centerline{\includegraphics[width=0.48\textwidth]{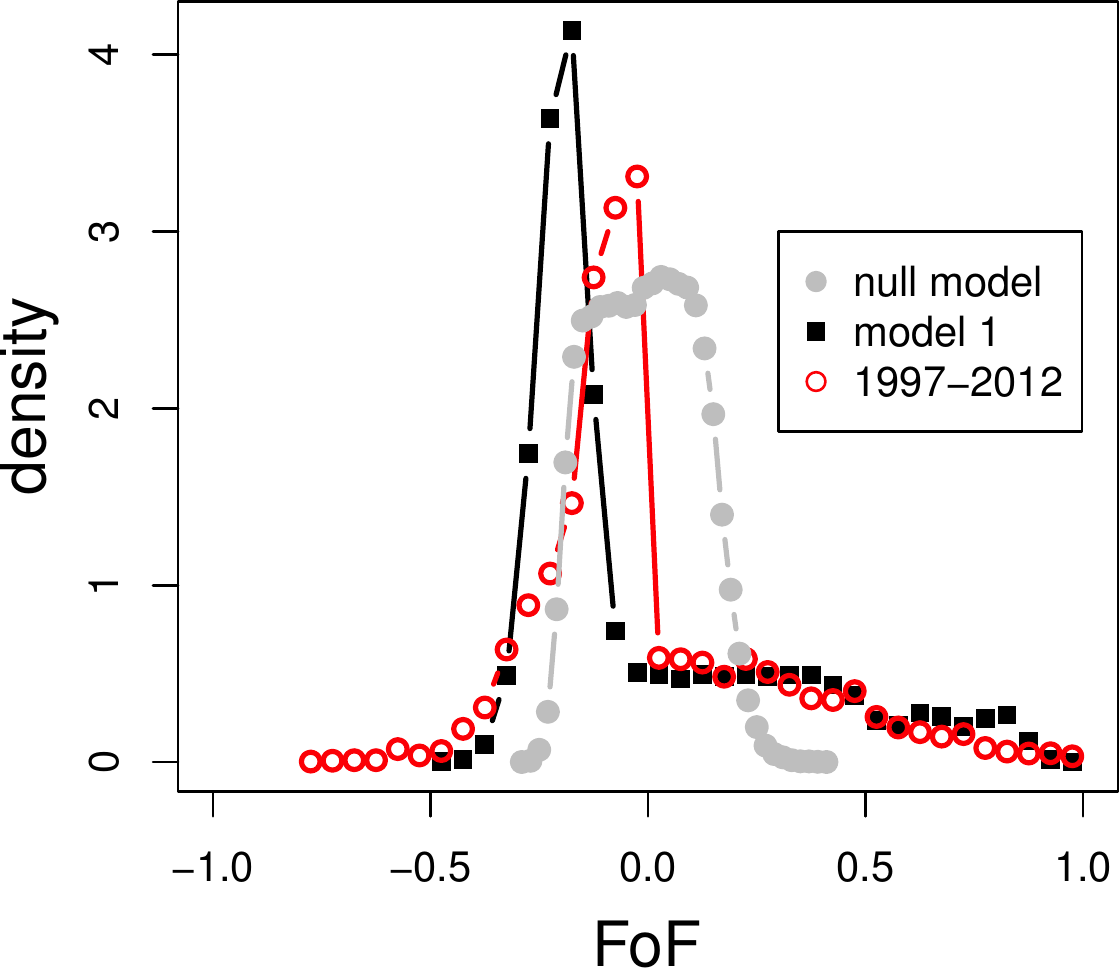}}
\vspace*{8pt}
\caption{Distribution of FoF values in contests between 1997 and 2012
  (red circles), compared with 10 simulations of the null model (gray
  points), and of model 1 (black squares).\label{fig:NullModel}}
\end{figure}

Figure \ref{fig:NullModel} shows the historical FoF distribution,
compared with the distributions from simulations of the null model and
model 1. It is easy to notice that the empirical distribution is
significantly different from the other two. We calculated goodness of
fit values for these differences, finding a $R^2 = 0.326$ and a
Kolmogorov-Smirnov statistic of $0.29$ for the difference between the
null model and the empirical data. For the case of model 1, the fit
improved to a $R^2 = 0.567$ and the KS statistic to $0.18$, but these
values allow us to conclude that the empirical FoF distribution is
significantly different from both models (KS test gave $p < 10^{-15}$
for both).

\subsection{A model for cultural affinity in Eurovision}

The above quantitative analysis reveals that the empirical FoF values
cannot be explained by random behavior under the contest rules.  In
the following, we propose a model to simulate Eurovision contests
under heterogeneous song quality and the influence of cultural
affinity, aiming at a better representation of the mechanism behind
the voting decisions of the participants.  

The model receives as an input a network in which nodes represent
countries, connected by edges with weights that represent cultural
affinities, i.e. a measure that takes high positive values for
culturally similar countries, and negative values for 
dissimilar cultures.  Thus, an edge $e_{v,c}$ connecting the node of
country $c_v$ to the node of country $c_c$ has a weight $w_{v,c}$ that
measures the cultural affinity of $c_v$ towards $c_c$.  This network
is composed of two subnetworks, with participant and only voting
countries.  The subnetwork between participants is fully connected,
directed, weighted network, and the only voting countries are
connected to all the participant ones by unidirectional weighted
links.

At the beginning of a simulation, we assign a quality value $q$ to
each participant, sampled uniformly at random from the distribution of
rescaled scores $s'_c$, shown in Figure \ref{fig:scoreDists}. This
value is an approximation of the artistic quality of a song
\cite{Ginsburgh2008}, which can be assumed to influence the final
outcome of the contest. Then the affinity model is simulated as
follows:

\fbox{
\parbox{0.85\textwidth}
 {\textbf{Affinity Model}\\
  scores = $[12,10,8,7,6,5,4,3,2,1]$\\
  w = affinity network\\
  For each competing country $c_c$:
  \begin{itemize}
  \item sample $q[c_c]$ from empirical $s'_c$ distribution
  \end{itemize}

  For each voting country $c_v$:
  \begin{enumerate}
  \item For each competing country $c_c$:
    \begin{itemize}
    \item assign $fit_v[c_c] = \alpha q[c_c] + (1-\alpha) w_{v,c}$
    \end{itemize}
  \item For each competing country $c_c$:
    \begin{itemize}
    \item If $rank(fit_v[c_c]) \leq 10$:  assign $p_{v,c} = scores[rank(fit_v[c_c])]$
    \item Else: assign $p_{v,c} = 0$
    \end{itemize}
  \end{enumerate}
}}

In the first step of the simulation, a country $c_v$ constructs a
local ranking of the other participant countries, by computing a value
$fit_v[c_c]$ that is a function of the weight $w_{v,c}$ of the edge
$e_{v,c}$, and the quality of the song, $q[c_c]$. We assume that this
function is a combination of both the quality and the affinity,
monotonically increasing with both. As an initial approximation, we
assume a linear combination of the form $\alpha q[c_c] + (1-\alpha)
w_{v,c}$, but future research can shed light on how countries combine
quality and cultural affinities when voting in Eurovision. This
function represents the combination of jury votes and televotes, as
empirical studies show that the jury is more influenced by the
artistic quality of a song than the televotes \cite{Haan2005}, which
seem to be driven by geographical and cultural biases. The current
rules of the contest give the same weight to both televotes and judge
votes, so we will choose $\alpha=0.5$ for our simulations. In the
second step, given the rank of each node, the agents cast their votes
in order, assigning them according to the voting scheme of Eurovision.

Under the lack of any other alternative assumption, we take edge
weights $w_{v,c}$ sampled from a normal distribution $N(\mu, \sigma)$
with parametrized values of the mean and standard deviation. We will
explore the role of these two values in reproducing the FoF
distribution. The output of the model is an artificial voting result
that can be compared with the real world data. In the following, we
present an analysis of the conditions under which this model provides
a more plausible FoF distribution, as compared to empirical data.

\subsection{Simulated FoF distributions}

To compare our model with the empirical FoF, we run a simulation
scheme that included all the combinations of values of $\mu$ in
$[-0.1,0.1]$ in increments of $0.01$ and $\sigma$ in $[0,0.1]$ in
increments of $0.005$. For each combination of parameter values, the
model was run 100 times, and the resulting FoF values were stored to
be compared with the empirical data.

To minimize computational efforts, we simulated a contest with 43
voting countries and 25 competing ones, which are values close to the
current editions of Eurovision. In our empirical validation, we focus
in the time range from 2004 until 2012, where the current final
structure was introduced. While the amounts of participants vary
within this period, their change is around 2 to 3 countries, allowing
us to use the same set of simulations to compare across years.

\begin{figure}[th]
\centerline{\includegraphics[width=0.9\textwidth]{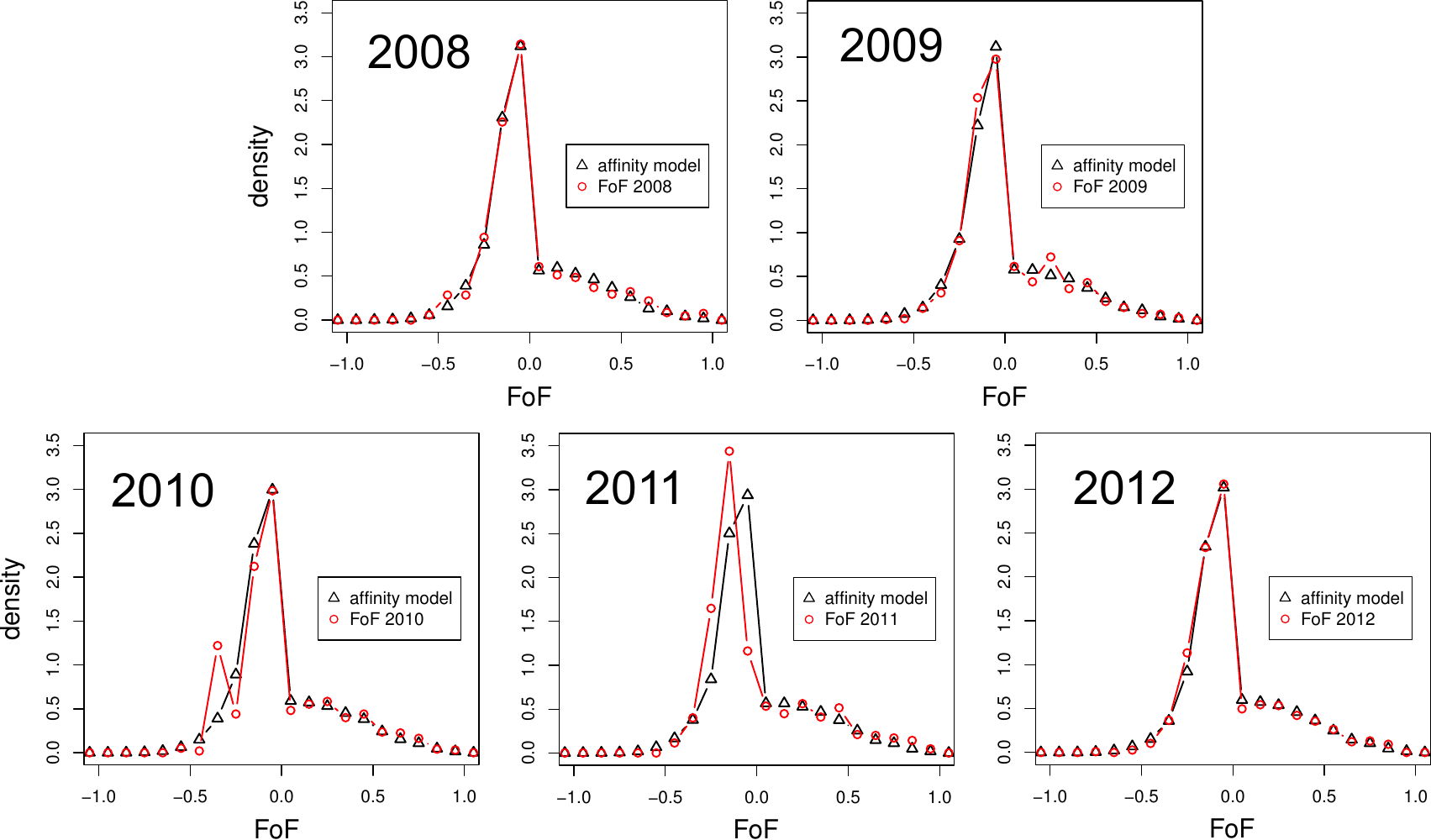}}
\vspace*{8pt}
\caption{ Distribution of empirical FoF values (red circles) between
  2008 and 2012, compared to best fits of the affinity model (black triangles).
  \label{fig:AffinityModelComparison}}
\end{figure}

After all simulations were run, we fitted the empirical FoF
distribution of each year against all the combinations of parameter
values. Our criterion to select the best fitting values was the
minimization of the Kolmogorov-Smirnov statistic \cite{Clauset2009}.
Figure \ref{fig:AffinityModelComparison} shows the model fit to the
last five editions of Eurovision. Generally, the model correctly
captures the shape of the distribution, with some discrepancies in
2010, where a second negative mode appears, and 2011, where the median
is shifted to the left. In Section \ref{sec:eu15} we provide a more
detailed analysis of the reasons that can explain these anomalies.

\begin{figure}[th]
\centerline{\includegraphics[width=0.7\textwidth]{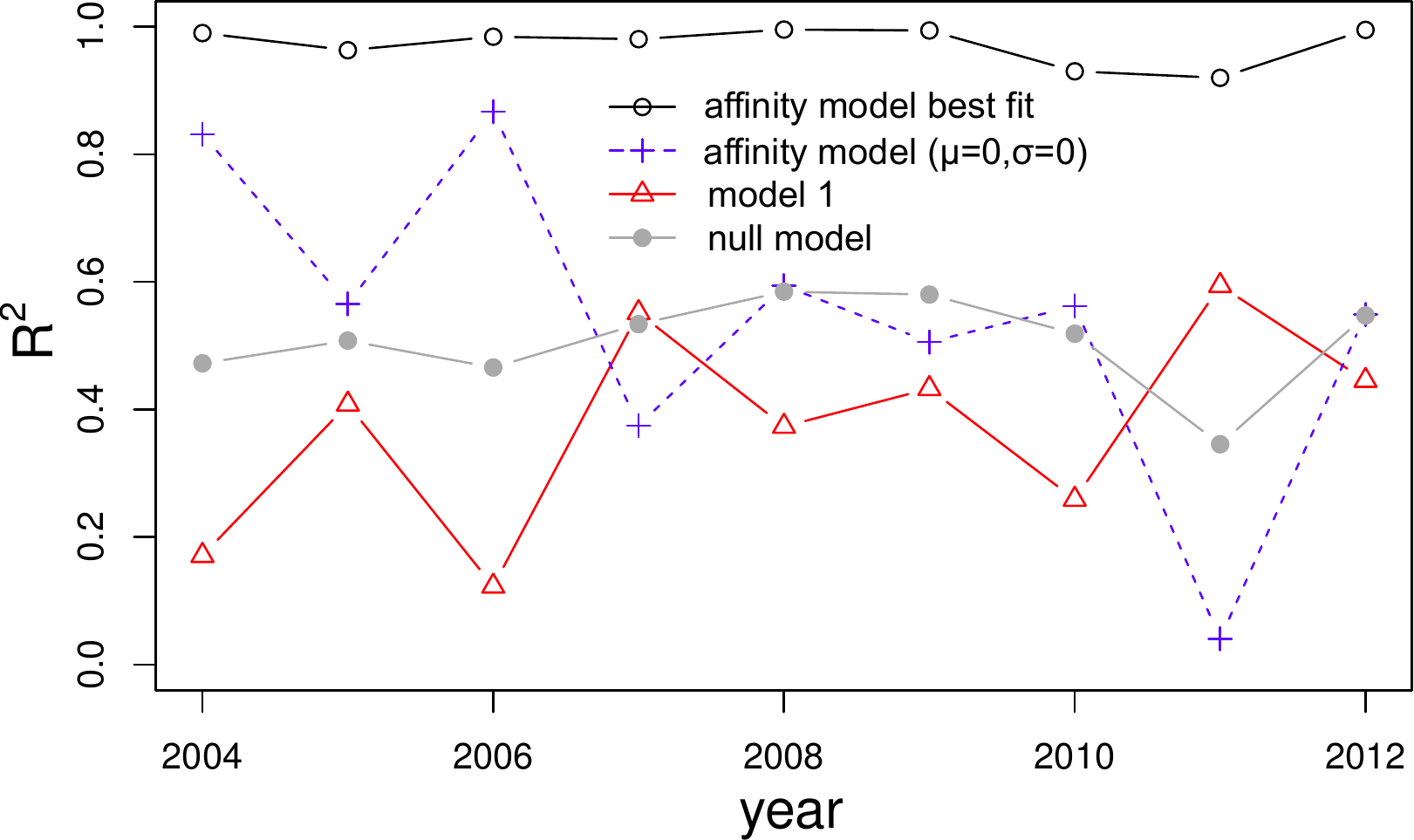}}
\vspace*{8pt}
\caption{$R^2$ values for model fits for the FoF distributions between
  2004 and 2012. The values reported for the affinity model correspond
  to the best fitting parameter values for the affinity distribution,
  and for $\mu=0$, $\sigma=0$.  \label{fig:AffinityModelR2}}
\end{figure}

Figure \ref{fig:AffinityModelR2} shows the $R^2$ values for the best
fits of the affinity model versus the FoF distributions of each year,
as well as the $R^2$ of the null model and model 1 as explained in
Section \ref{sec:nullmodels}. The goodness of fit of the affinity
model is above the other two for all years, typically explaining an
additional 40\% of the variance of the FoF distribution.  While the
$R^2$ shows that the affinity distribution plays an important role,
this affinity model is still insufficient to capture all the dynamics
that produce the shape of the FoF distribution. The results of the
Kolmogorov-Smirnov tests gave distances between 0.06 and 0.12, with p
values below $10^{-5}$ for each year. This test shows that we can
reject the hypothesis that simulated and empirical FoF come from the
same distribution. Nevertheless, the better fit of the affinity model
in comparison with the null model and model 1 allows us to validate
that the FoF coefficient reflects the network of affinity between
countries, yet the model definition can be improved.

The values of $\mu$ and $\sigma$ for the best fit of each year contain
information on what is the most likely distribution of cultural
affinities. For each year, the best fitting $\sigma$ have the same
value of $0.075$, while the best fitting $\mu$ span equally across
positive and negative values. This latter fact is not surprising, as
the rank transformation of the affinity model destroys the influence
of the mean of the affinity distribution. We can say that, in the
range of explored values, $\mu$ acts as a free parameter, and $\sigma$
as a constant.

Among the parameter values of our simulations, an interesting case is
$\mu=0$, $\sigma=0$, which corresponds to the deterministic case in
which song quality is the only criterion for choosing the votes.  In
that case, cultural affinity does not influence the simulated votes,
and only song heterogeneity and contest rules account for the final
FoF values. The quality of the fit in that case is also shown in
Figure \ref{fig:AffinityModelR2}, revealing the inconsistent quality
of ignoring affinity as a factor in the model simulation. From these
results we conclude that the Friend-or-Foe coefficient indeed contains
significant information about the affinities among participant
countries, and its distribution of values cannot be explained by the
contest rules alone.

\section{Empirical analysis of FoF networks}

\subsection{Votes and FoF networks}

The above numerical analysis draws a connection between cultural
affinities and the Friend-or-Foe coefficient. When analyzing empirical
Eurovision data, it follows to ask whether the FoF can be used to
reveal patterns beyond those that can be found analyzing voting scores
alone, which is the standard technique used in previous research
\cite{Yair1995,Dekker2007}. This traditional approach models the
result of a Eurovision contest as a network in which nodes represent
participating countries, and directed edges connect nodes with weights
according to the points assigned in the contest. We define a new type
of network with the same nodes, but fully connected with directed
edges $e(v,c)$ with signed weights corresponding to $FoF(c_v, c_c)$.

\begin{figure}[th]
\centerline{
\includegraphics[width=0.46\textwidth]{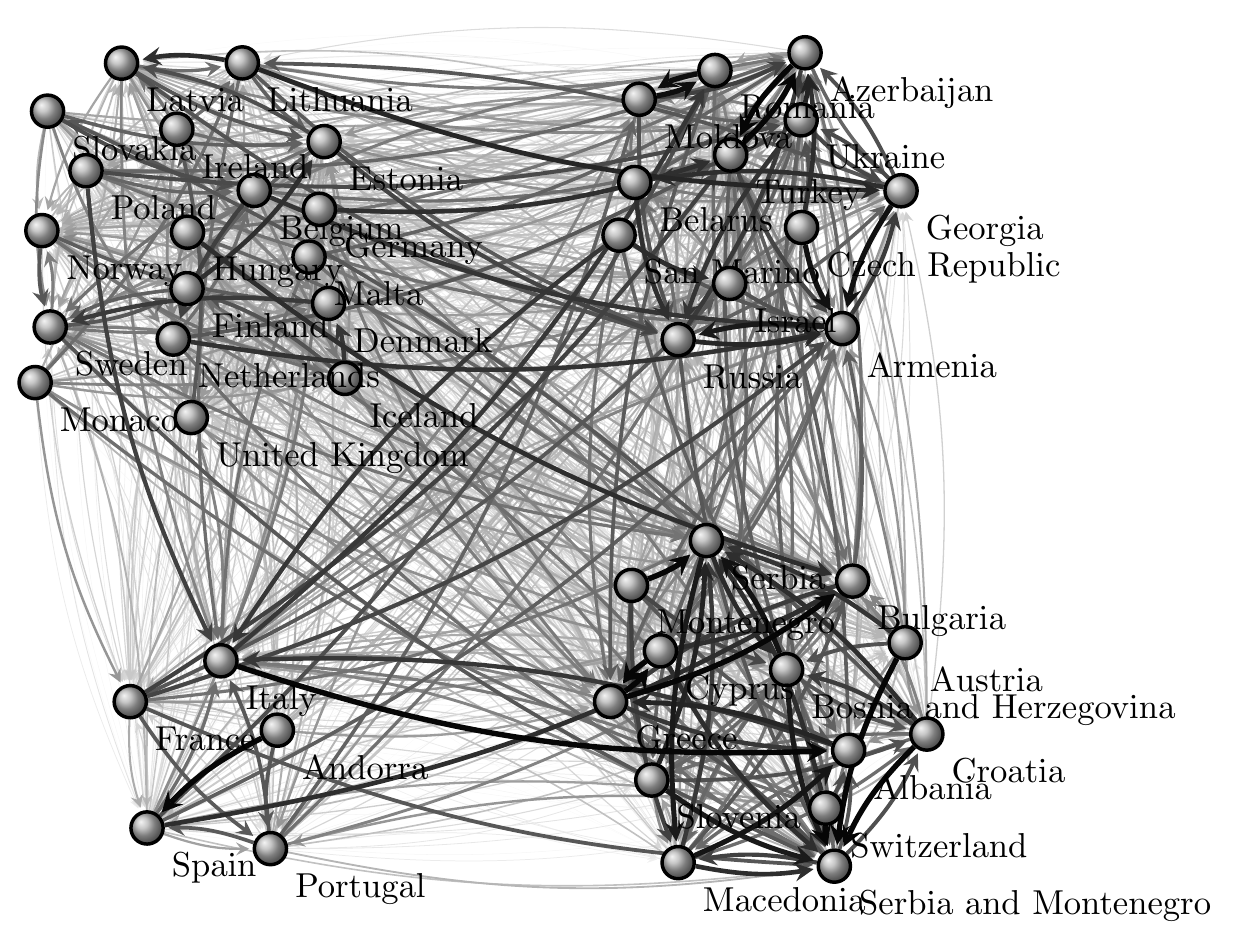}\hfill
\includegraphics[width=0.53\textwidth]{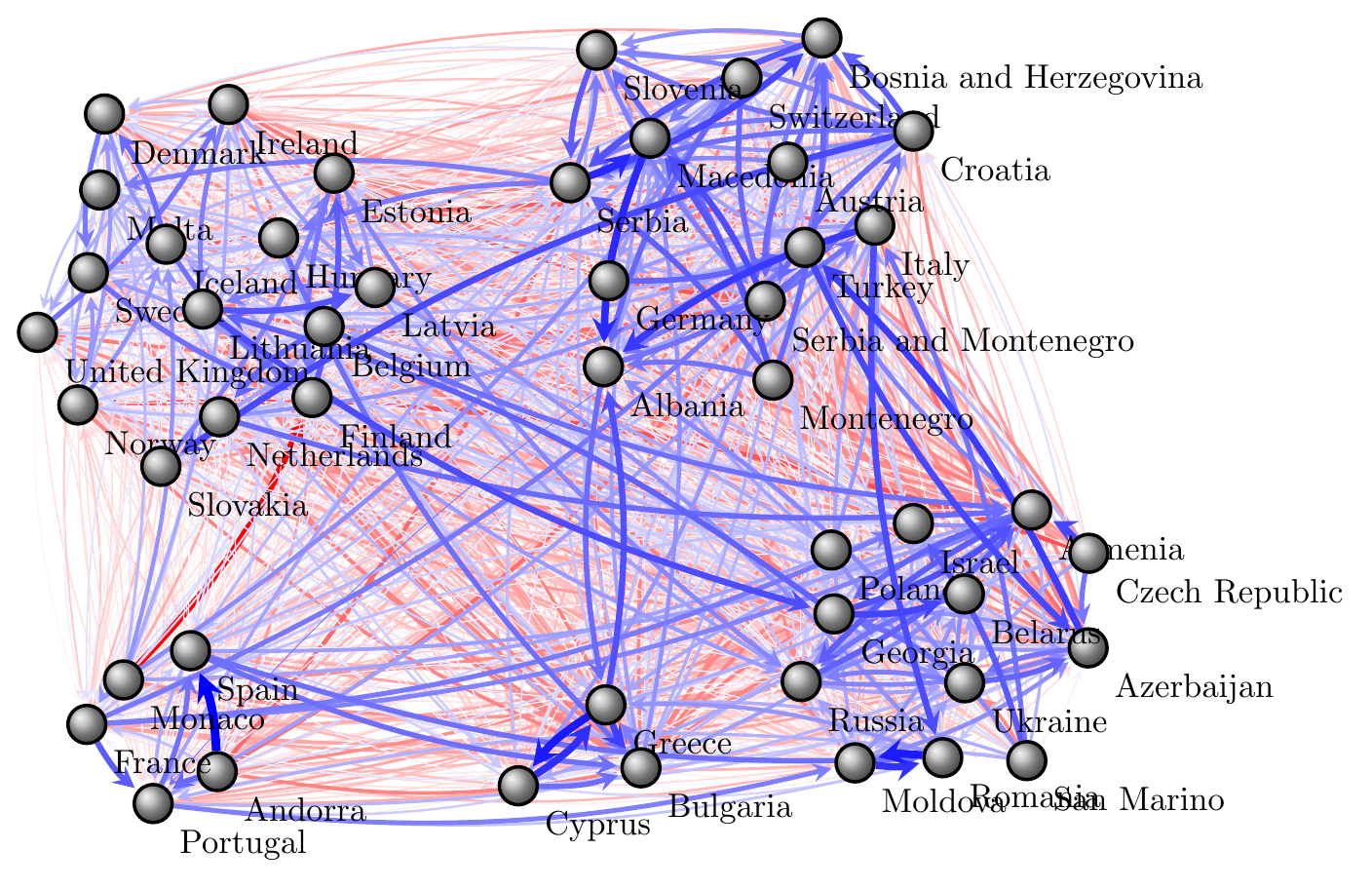}}
\vspace*{8pt}
\caption{Left: mean score network between countries for the period
  1997-2012. Edge darkness and widths are proportional to the average
  score from one country to another, and nodes are located in four
  communities according to maximal weighted modularity. Right: mean
  FoF network for the same period, displaying negative FoF in red, and
  positive Fof in blue. Edge width and darkness are proportional to
  the absolute value of FoF. Nodes are arranged in five communities
  according to maximal weighted, signed
  modularity. \label{fig:Clustered-networks}}
\end{figure}

We explore the time aggregation of these networks by averaging the
scores and FoF between each pair of countries that has participated in
Eurovision. We focus on the time interval between 1997 and 2012,
capturing the contest results since televoting was introduced.  This
leads to two different networks: a \emph{mean score network} with mean
scores as edge weights, and a \emph{mean FoF network} with mean
Friend-or-Foe coefficients as weights. These two networks are shown in
Figure \ref{fig:Clustered-networks}, visualized with the the
Cuttlefish Network Workbench
\footnote{\url{www.cuttlefish.sourceforge.net}}.

A common research question on \emph{eurovipsephology} is the analysis
of clusters and subcommunities among participant countries. A variety
of data transformations and techniques have been applied, including
dimensionality reduction of reciprocal structures \cite{Yair1996} and
correlations \cite{Fenn2006}, numerical comparisons at certain
significance levels \cite{Dekker2007,Gatherer2006}, clusters created
by discarding votes below varying thresholds
\cite{Dekker2007,Saavedra2007}, and techniques to find overlapping
voting clusters \cite{Orgaz2011}. The current state of the art in
community detection is based on Q-modularity \cite{Newman2004}, which
has previously been applied to assess the quality of threshold-based
dendrograms in Eurovision \cite{Saavedra2007}. In the following, we
define and apply two metrics to quantify and compare the modularity of
the votes and FoF networks.

\subsection{Modularity metrics}

\label{sec:modularity}

The modularity of the mean score network can be measured with respect
to a partition of the nodes $C_i$, which assigns a community value
to each node $i$. This partition is quantified through the function
$\delta(C_v, C_c)$, which has value $1$ if $v$ and $c$ belong to the
same community, and $0$ otherwise. In this setup, we can compare the
modularity of a partition of the network with the random case, by
means of the following equation \cite{Newman2004}:
\begin{equation}
Q_s = \frac{1}{2S} \sum_v \sum_c \left ( \overline{p}_{v,c}   - \frac{p_v^{out}p_c^{in}}{2S}        \right )
\delta(C_v, C_c) 
\label{eq:weightedmodularity}
\end{equation}
where $\overline{p}_{v,c}$ is the mean score that $c_v$ has given to
$c_c$, $p_v^{out}= \sum_i \overline{p}_{v,i}$, $p_c^{in}= \sum_i
\overline{p}_{i,c}=s_c$ and $S$ is the total amount of points given in
the contest.  Since the contest rules were fixed in 1975,
$p_v^{out}=58$, and $S=58N_v$, but the problem of finding the best
community structure is still NP-hard, requiring approximated
algorithms when networks are not very small.

In our mean score network, the weights of the links are assigned as
the mean score a country gives to another, calculated over the 16
contests between 1997 and 2012. This way we do not set any ad hoc
threshold, and we account for all the data available since televoting
was introduced in the contest. We computed optimal communities through
10000 bootstrapped heuristic searches \cite{Arenas2007}, finding four
communities with a modularity of $0.166$. The layout of the left
network of Figure \ref{fig:Clustered-networks} shows these four
differentiated subcommunities.

The mean FoF network differs from the mean score network in the fact
that it contains signed weights, as mean FoF values can be
negative. To make use of this feature, we use the definition of
signed, weighted modularity \cite{Gomez2009}:
\begin{equation}
Q_{fof} = \frac{1}{2f^+ + 2f^-} \sum_v \sum_c \left [ \overline{Fof}(c_v,c_c)  -
  \left(  \frac{f_v^{+,out}f_c^{+,in}}{2f^+} - \frac{f_v^{-,out}f_c^{-,in}}{2f^-}     \right) \right  ] \delta(C_v,C_c)  
\label{eq:signedmodularity}
\end{equation}

where $f^{\pm} $ are the total sums of positive and negative FoF
values, and $f_c^{\pm, in/out}$ are the sums of incoming and outgoing
positive and negative FoF values for country $c_c$. The rationale
behind Equation \ref{eq:signedmodularity} is to measure the density of
positive and negative FoF inside the communities, compared with the
random case of an uncorrelated network. $Q_{fof}$ will increase when
the communities contain more positive FoF, as well as when the
negative FoF are kept across communities.

We run over the mean Fof network the same method as for the mean score
network, looking for partitions that maximize the internal positive
FoF of the communities, while minimizing the amount of internal
negative FoF at the same time. The right network of Figure
\ref{fig:Clustered-networks} shows the five clusters we found, having
$Q_{fof}= 0.252$. This higher modularity of the mean FoF network,
in comparison with the mean score network, reveals the added value of
the Friend-or-Foe coefficient. We found a network partition that
differs more from the random case, as compared with the mean score
network, as the Q-modularity is lower when ignoring the final result
of the contest.

\subsection{Dynamics of modularity and polarization}
\label{sec:moddyn}
While above modularity metrics highlight subcommunities, other
patterns might arise from the results of Eurovision contests. From a
macroscopic point of view, FoF values can reveal different levels of
strength in the biases of country votes, without assuming any
particular division in communities. To measure this overall level of
``disagreement'' across participants, we define the polarization of
FoF as:
\begin{equation}
  Pol(t) = \sqrt{\frac{1}{E_t}\sum_{c,v} (FoF_t(c_c,c_v) - \langle FoF_t \rangle)^2}
\label{eq:polarization}
\end{equation}
which is essentially the standard deviation of FoF across all the
pairs of countries in the network, which amount to $E_t$.  This
polarization metric takes higher values when voting biases are strong,
in comparison to the artistic component of the contest. If all
countries agreed on the best songs in the same manner, the
polarization would have a value close to zero.

The definition of polarization of Equation \ref{eq:polarization}
allows us to track the overall strength of individual biases in the
history of Eurovision. Similarly, we extend the definitions of
modularity explained above, aiming to measure the modularity in the
scores network $Q_s(t)$ and the FoF network $Q_{fof}(t)$ of each year
$t$. These two time series of modularities are calculated as Equations
\ref{eq:weightedmodularity} and \ref{eq:signedmodularity},
substituting average values with the instances of each year
,$p_{v,c}(t)$, and $FoF_t(c_c, c_v)$.

\begin{figure}[th]
\centerline{\includegraphics[width=0.75\textwidth]{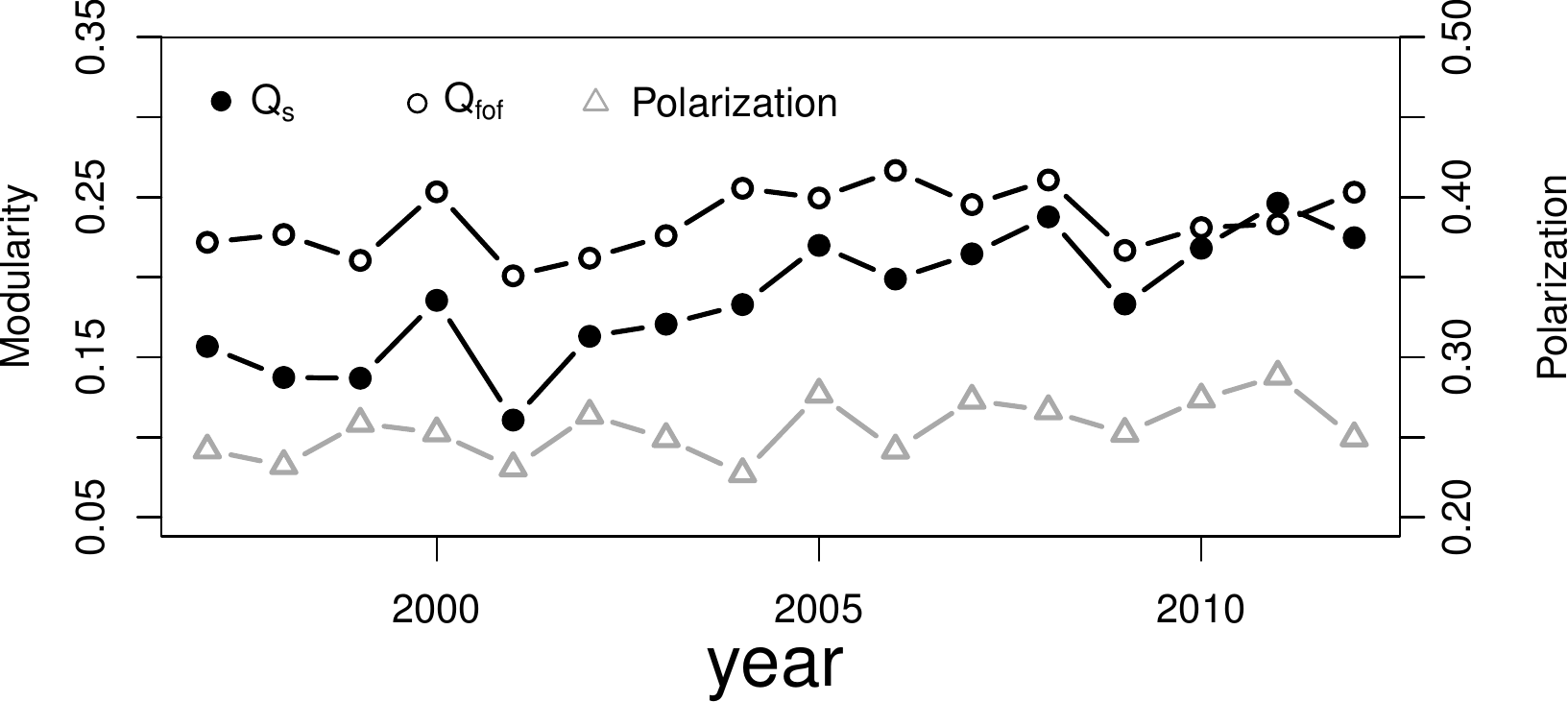}}
\vspace*{8pt}
\caption{Time series of votes modularity $Q_s$ (black points), FoF
  modularity $Q_{fof}$ (circles), and polarization (gray triangles).
  \label{fig:Modularity-year}}
\end{figure}

Figure \ref{fig:Modularity-year} shows the time series of both
modularities and polarization. Our first observation is an increasing
pattern of $Q_s$, having always values below $Q_{fof}$, with the
exception of 2011. As mentioned in Section \ref{sec:controversies},
the organization of the Eurovision song contest has been accused of
having an increasing level of unfairness and lack of artistic content.
This is in line with the increasing pattern of $Q_s$, which we
validated by measuring the correlation between $Q_s$ and $t$, which
has a value of $0.817$ (p-value= $0.0001097$).

The score modularity approaches the FoF modularity from below, which
seems to keep constant through the ovserved time period. Testing this,
we computed the correlation between $Q_{fof}(t)$ and $t$ finding a
value of $0.402$ with low significance (p-value = $0.1224$). This lack
of a linear trend in $Q_{fof}$ shows how the FoF reveal rather stable
patterns of European culture, as opposed with the increasing pattern
of modularity in the scores.

We observe a rather stable level of polarization, which we test by
calculating the correlation between $Pol(t)$ and $t$. This
calculation gives a value of $0.521$ under a significance level that
very arguable could reject the null hypothesis (p-value =
$0.03816$). Therefore, we refrain from concluding that the
polarization of FoF shows any linear time pattern of generalized
biases, when looking at all the countries participating in
Eurovision. In the following section, we focus on a subset of the
countries participating in Eurovision, testing the relation between
their economic situation and the FoF polarization among them.

\section{Cultural affinity and economy in the EU}
\label{sec:eu15}

\subsection{The EU-15 subnetwork}

One of the main motivations for studying Eurovision data is its
representativeness of the whole Europe, giving the possibility to
study the relations among countries. It is particularly relevant to
explore its relation to the events of the European debt crisis,
testing if there is a relation between cultural affinity and the
economy at large.  In this section, we focus on a subset of countries
called the EU-15, which is the set of members of the European Union
since 1995.

\begin{figure}[th]
  \centerline{\hfill 2007\hfill \hfill 2008 \hfill \hfill 2009 \hfill}
  \centerline{\includegraphics[width=0.33\textwidth]{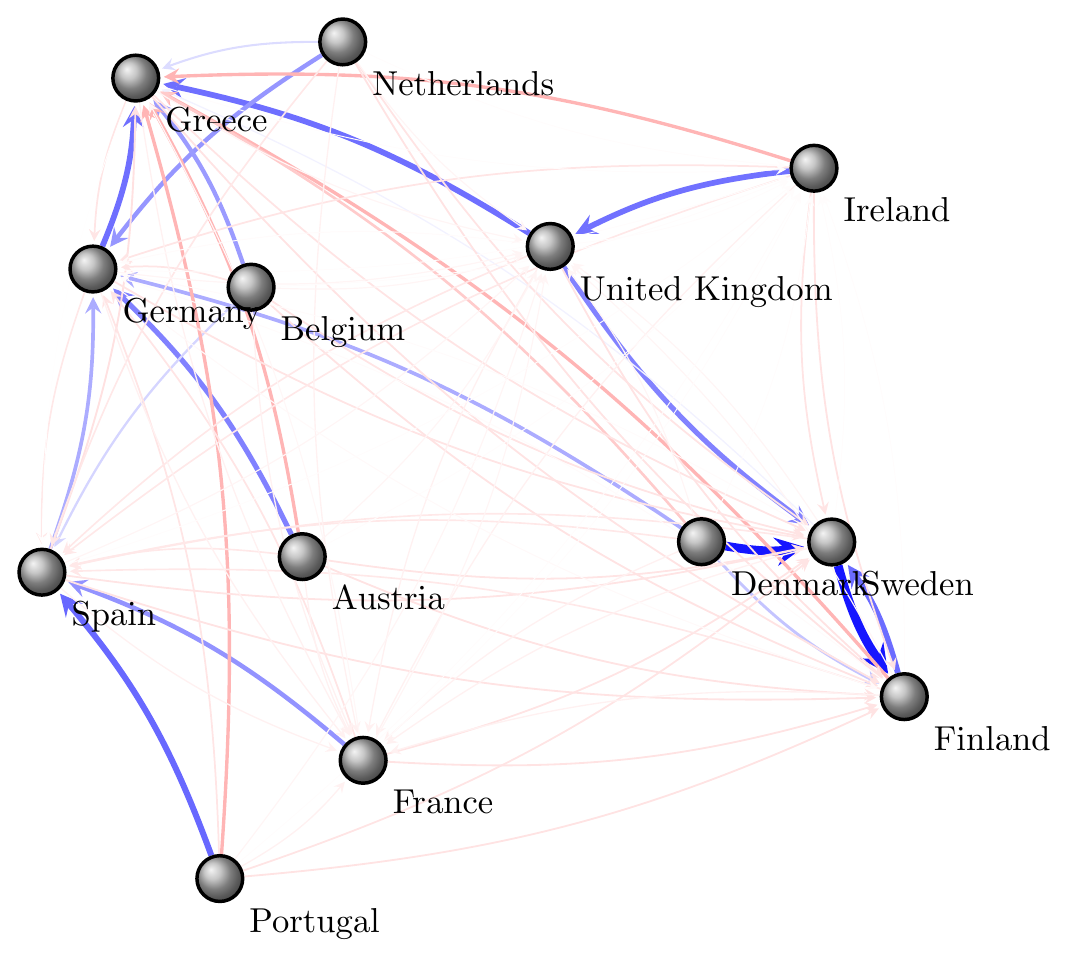}
  \hfill \includegraphics[width=0.33\textwidth]{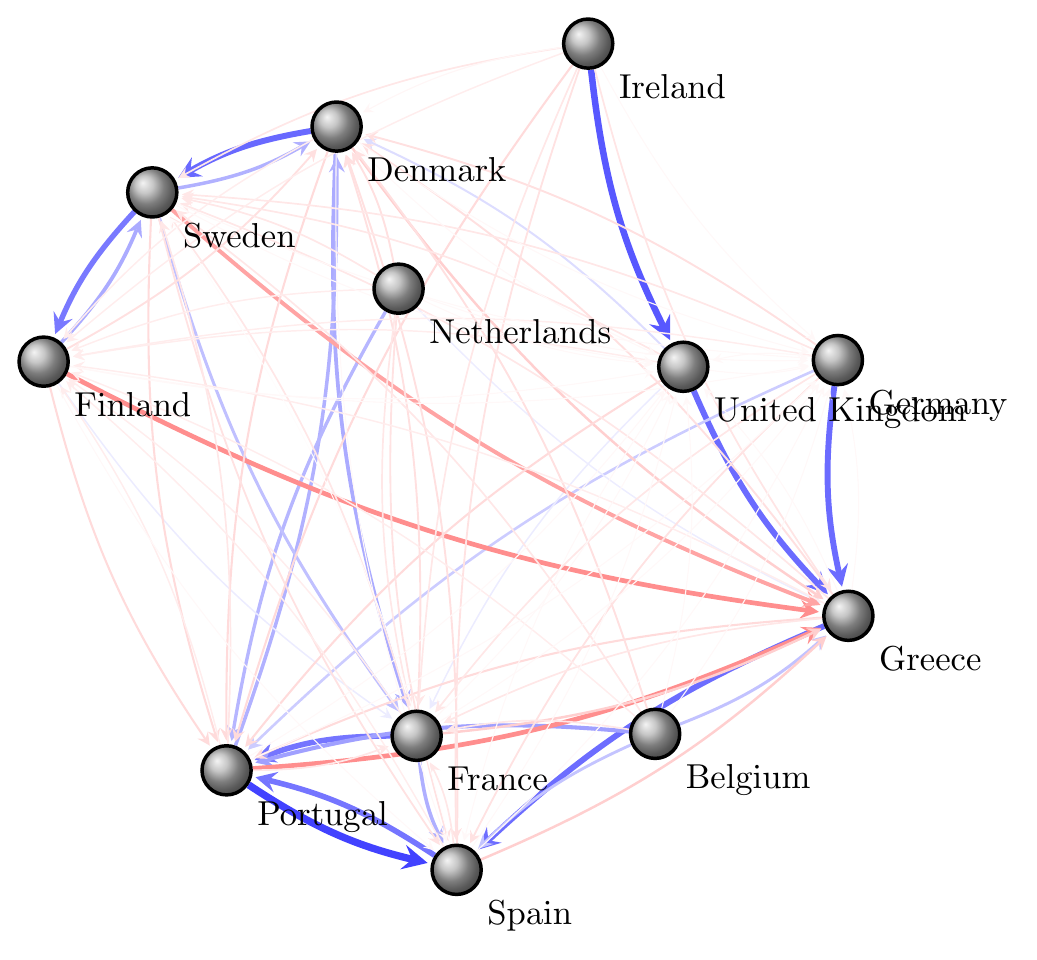}
 \hfill \includegraphics[width=0.33\textwidth]{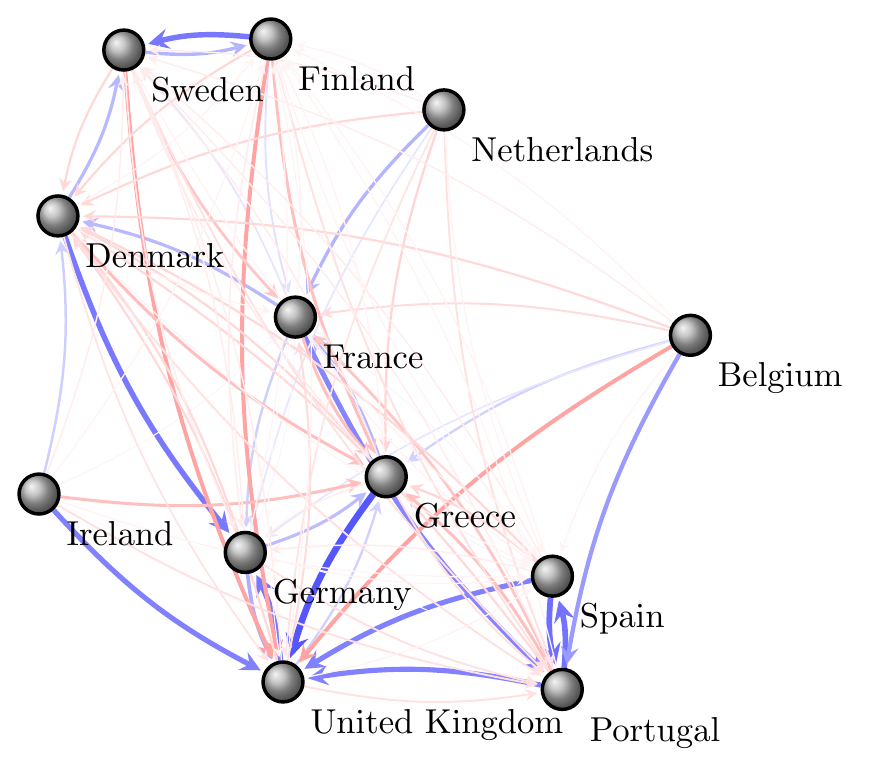}}
  \centerline{\hfill 2010\hfill \hfill 2011 \hfill \hfill 2012 \hfill}
  \centerline{\includegraphics[width=0.33\textwidth]{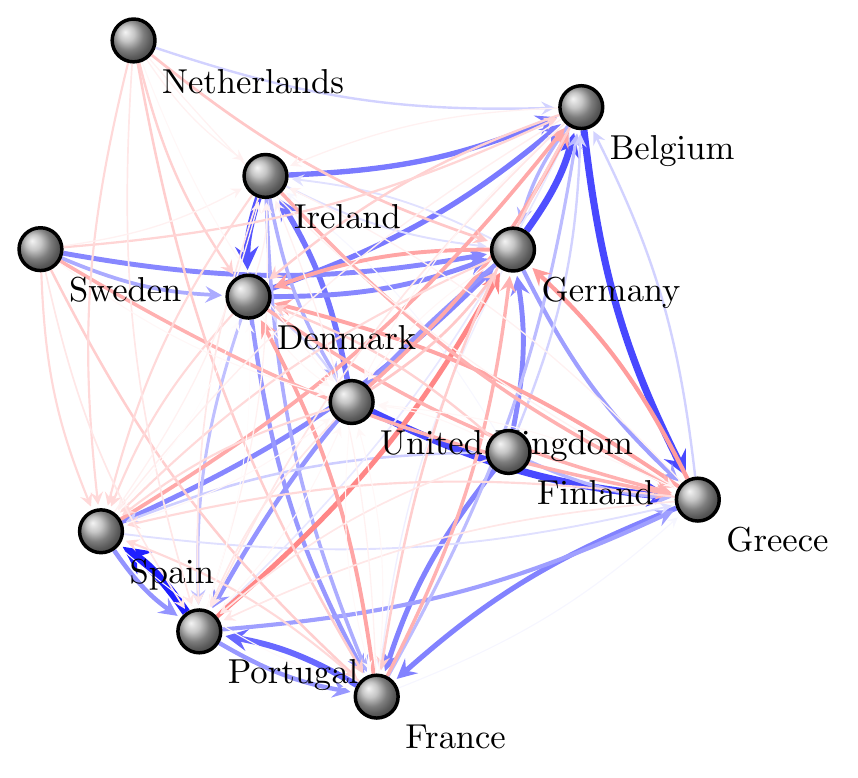}
  \hfill \includegraphics[width=0.33\textwidth]{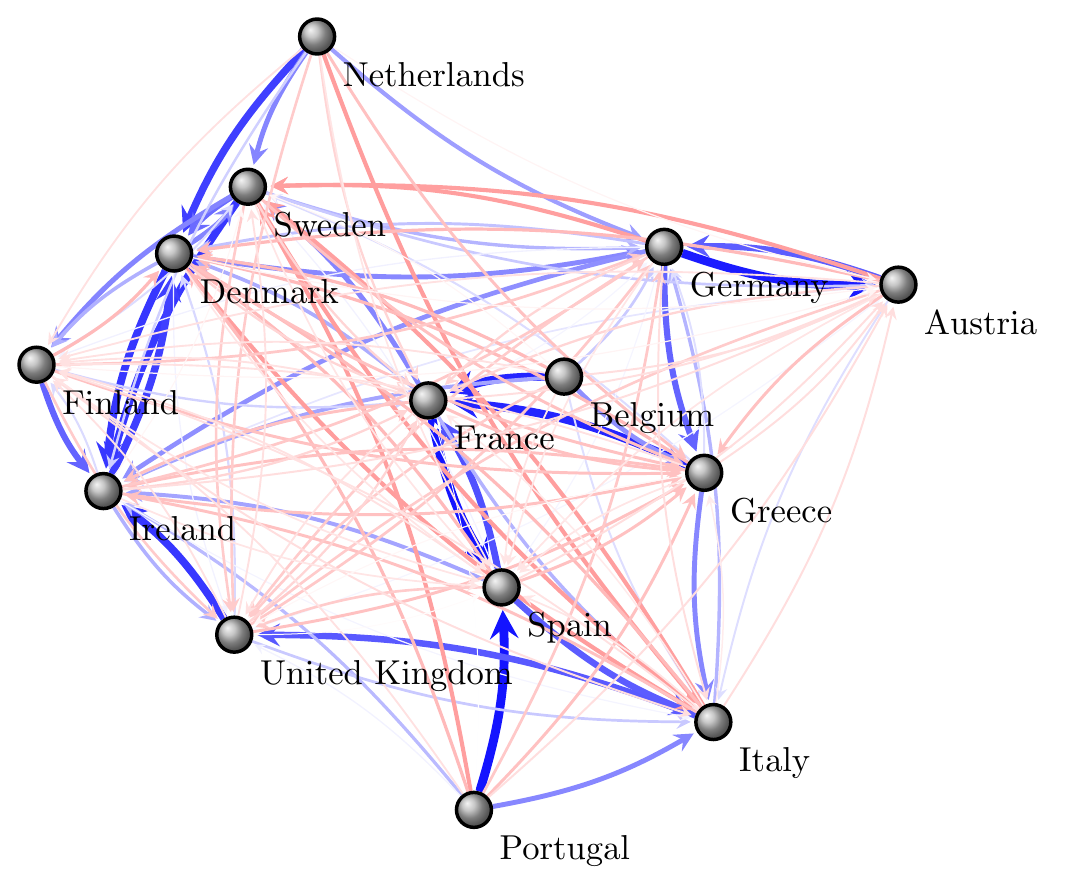}
 \hfill \includegraphics[width=0.33\textwidth]{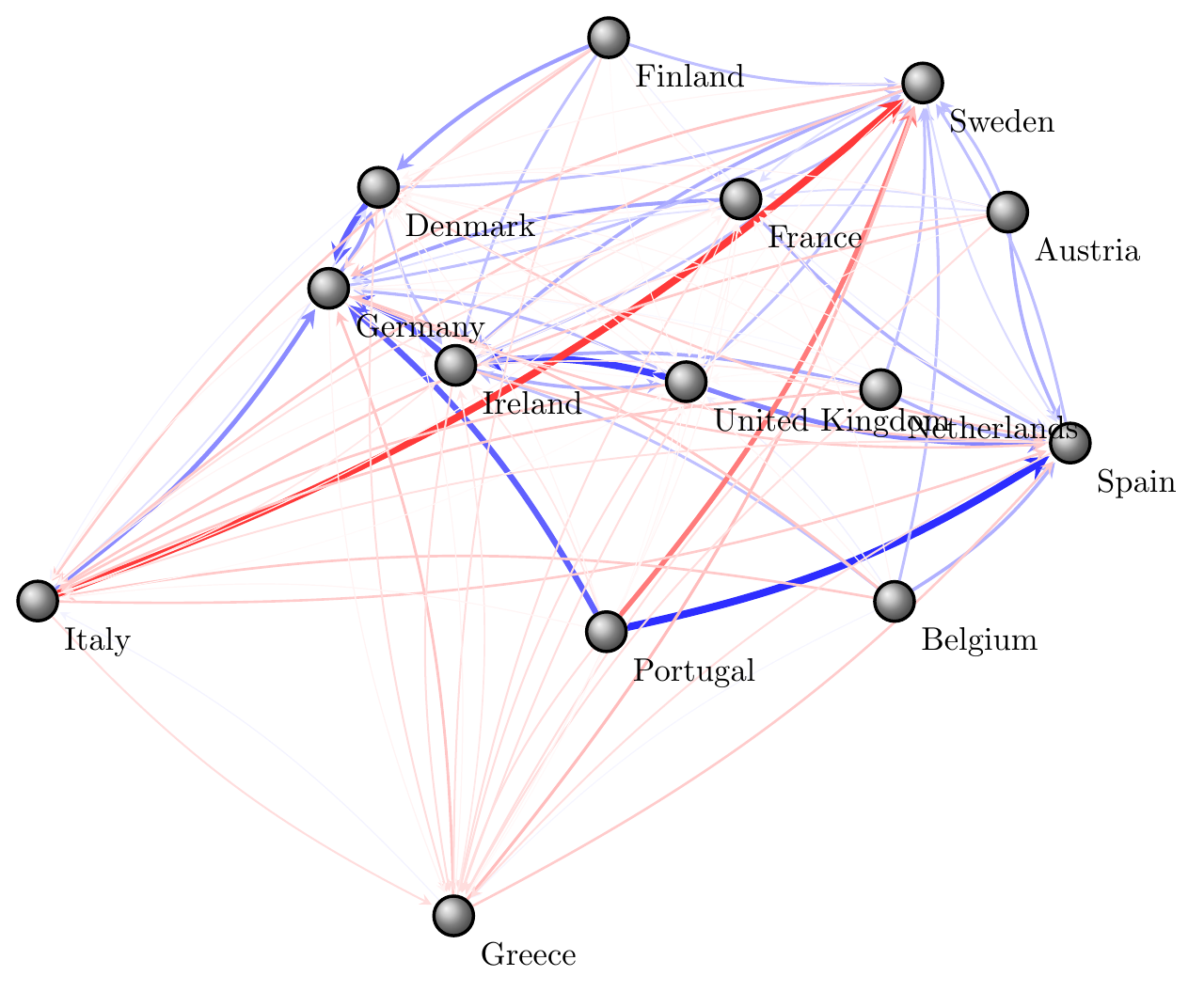}}
\vspace*{8pt}
\caption{ EU-15 subnetworks between 2007 and 2012. Each node
  represents a EU-15 country taking part in Eurovision that year. Each
  directed edge has a darkness and width proportional to the absolute
  value of the FoF coefficient from one country to another, colored
  blue for positive FoF and red for negative FoF.\label{fig:EU15nets}}
\end{figure}

Figure \ref{fig:EU15nets} shows the EU-15 FoF subnetworks for the
contest editions between 2007 and 2012. Blue edges represent positive
FoF with width and darkness proportional to the FoF of one country
towards the other. Red edges represent negative FoF in the same
manner, with darker and wider edges for more negative FoF.  These
networks show that in the years 2010 and 2011 the overall width of
both red and blue edges is stronger than for the rest of the years. On
the other hand, the network of 2007 seems to be more divided in small
clusters of positive FoF, while the later ones seem to form larger
clusters with lots of negative and positive FoF. In the following, we
measure modularity and polarization metrics for these networks, in
order to quantify these observations.

\subsection{Exact modularity and polarization}

For the whole Eurovision network, we used heuristic methods to find
optimal communities with maximal modularity (Section
\ref{sec:moddyn}). The case of the EU-15 subnetwork is much smaller,
with 12 to 14 participating countries each year, allowing us to apply
exhaustive search. For each edition of Eurovision between 1997 and
2012, we enumerated all the possible partitions of nodes, and
computed both $Q_s(t)$ and $Q_{fof}(t)$, finding the global maxima of
both. This required a considerable amount of computing power, as the
amount of partitions of a network of 14 nodes is the 14th Bell number,
which has an order of magnitude of $10^8$.

\begin{figure}[th]
\centerline{\includegraphics[width=0.75\textwidth]{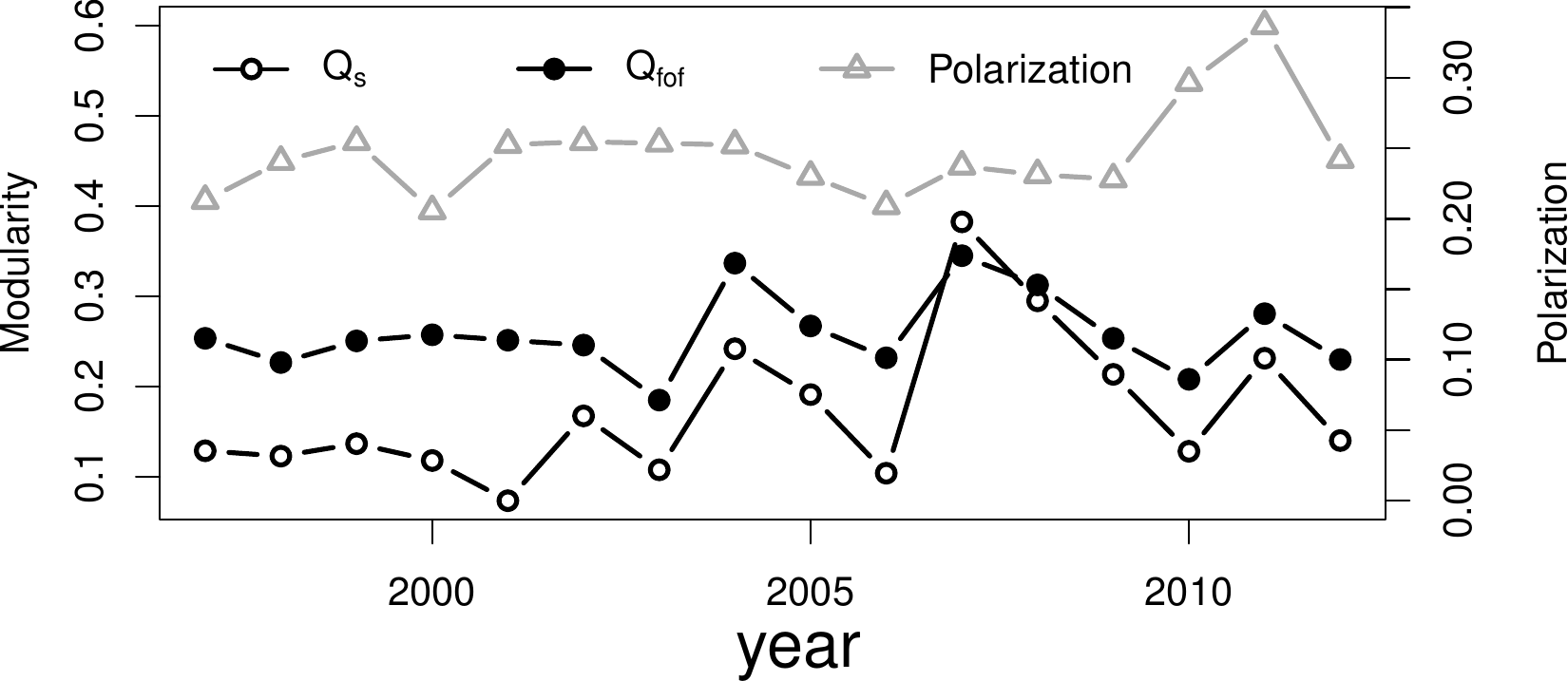}}
\vspace*{8pt}
\caption{Time series of votes modularity $Q_s$ (black points), FoF
  modularity $Q_{fof}$ (circles), and polarization (gray triangles)
  for the EU-15 subnetwork.
  \label{fig:Modularity-EU15}}
\end{figure}

The time series of the exact modularities and polarization in the
EU-15 are shown in Figure \ref{fig:Modularity-EU15}. Similarly to the
whole Eurovision network, the scores modularity keeps below the FoF
modularity, only having a slightly higher value during 2007. The peak
in 2007 validates our observation over Figure \ref{fig:EU15nets}, that
in 2007 the countries of the EU-15 could be divided in some tightly
connected subcommunities. In addition, the EU-15 subnetwork does not
show an increasing trend, neither of score modularity ($\rho(Q_s,
t)=0.421$, p-value = $0.1038$), FoF modularity ($\rho(Q_{fof},
t)=0.137$, p-value= $0.6132$), nor polarization ($\rho(Pol(t),t)=0.42$,
p-value = $0.106$).

\subsection{Relation between polarization and debt}

A careful observation of the time series of polarization in the EU-15
subnetwork reveals a peaked value in 2010 and 2011, coinciding with
the loans and austerity measures in Portugal, Ireland, Italy, Greece,
and Spain. As a comparison, the polarization keeps relatively stable
between 1997 and 2009, leading us to formulate the hypothesis that the
polarization in the EU-15 subnetwork is related to the European debt
crisis. To empirically test this hypothesis, we need a quantitative
indicator of the state of the European economy. For this purpose, we
use the interest rate of the sovereign bonds of the countries of the
Eurozone, which is commonly discussed as a method to assess the state
of the European
economy\footnote{\url{http://en.wikipedia.org/wiki/European_sovereign-debt_crisis}}.
In our analysis, we use the mean interest rate of the long-term
sovereign bonds of the 12 Euro founder countries, all part of the
EU-15\footnote{Belgium, Germany, Ireland, Greece, Spain, France,
  Italy, Luxembourg, Netherlands, Austria, Portugal, and
  Finland.}. The dataset we used is based on harmonized combinations
of primary and secondary markets, and is provided by the European
Central Bank
\footnote{\url{http://www.ecb.int/stats/money/long/html/index.en.html}}.
We used this mean interest rate as a standard metric to quantify the
evolution of the EU debt crisis.

\begin{figure}[th]
\centerline{\includegraphics[width=0.85\textwidth]{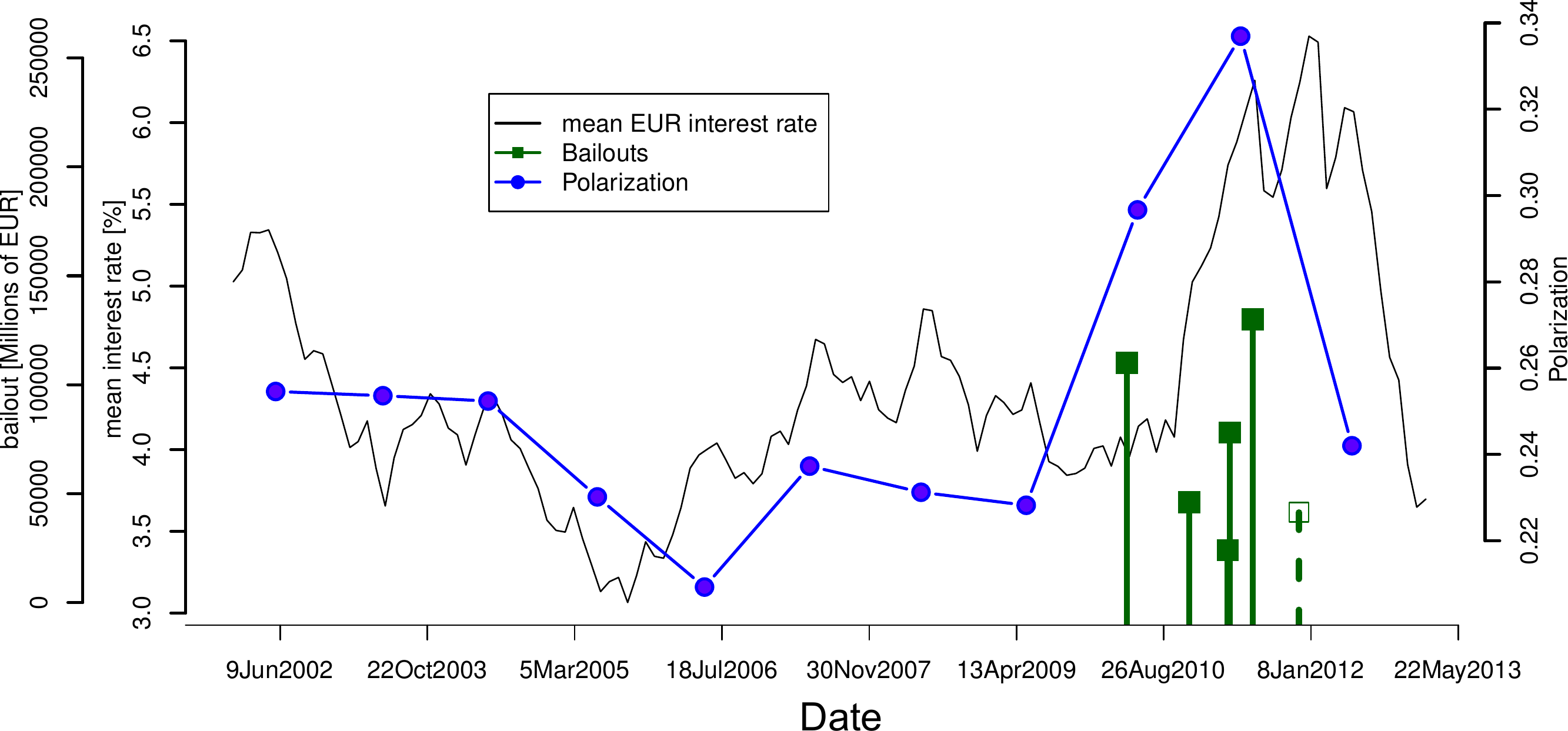}}
\vspace*{8pt}
\caption{Time series of EU-15 polarization (blue dots) and monthly
  mean sovereign bond interest rate for Euro-funding countries (black
  line). Green squares indicates dates and volumes of bailout loans to
  EU countries (Greece, Portugal, Ireland) until January 2013. The
  last green square indicates the bailout to the Spanish private
  banking sector. \label{fig:Debt-Pol-TS}}
\end{figure}

\begin{figure}[th]
\centerline{
\includegraphics[width=0.48\textwidth]{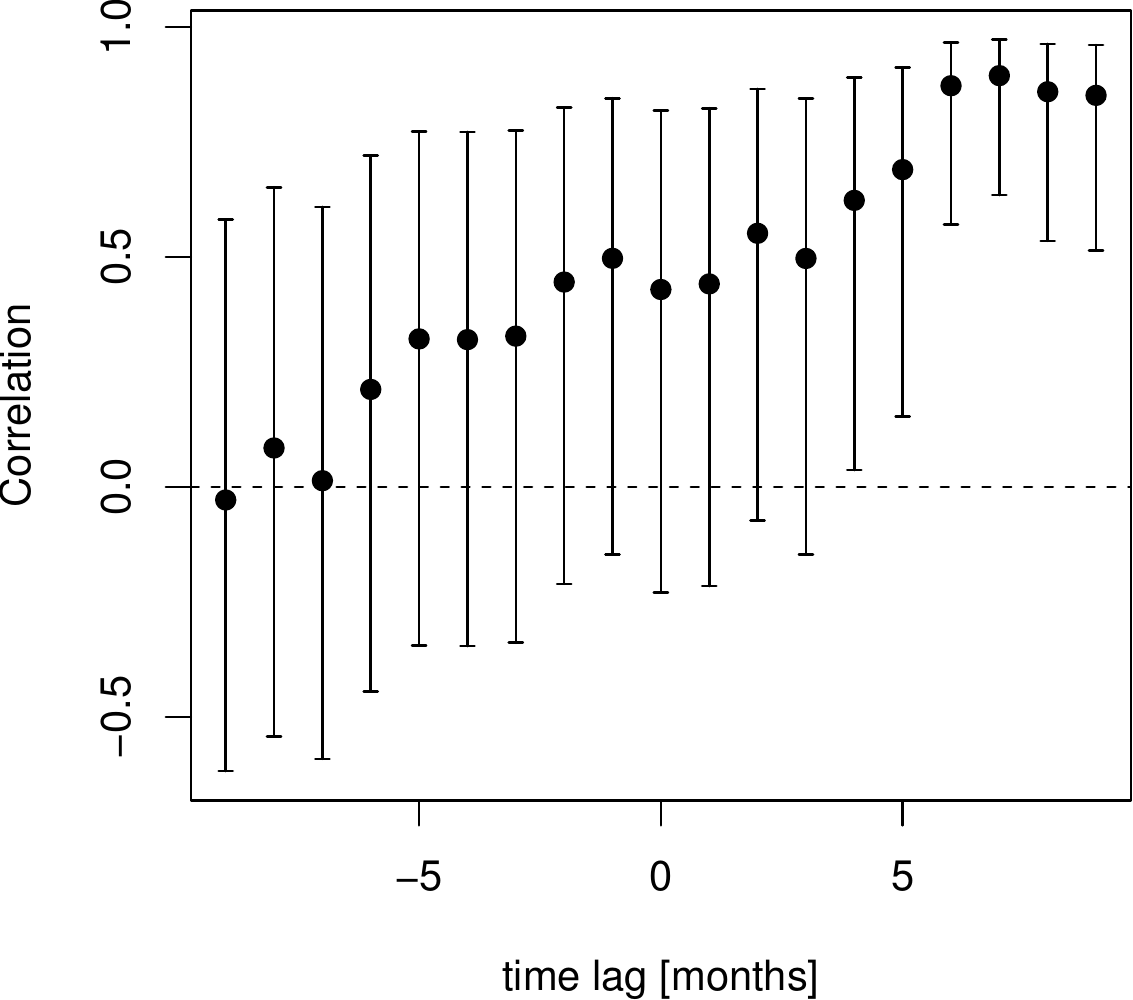}
\hfill \includegraphics[width=0.5\textwidth]{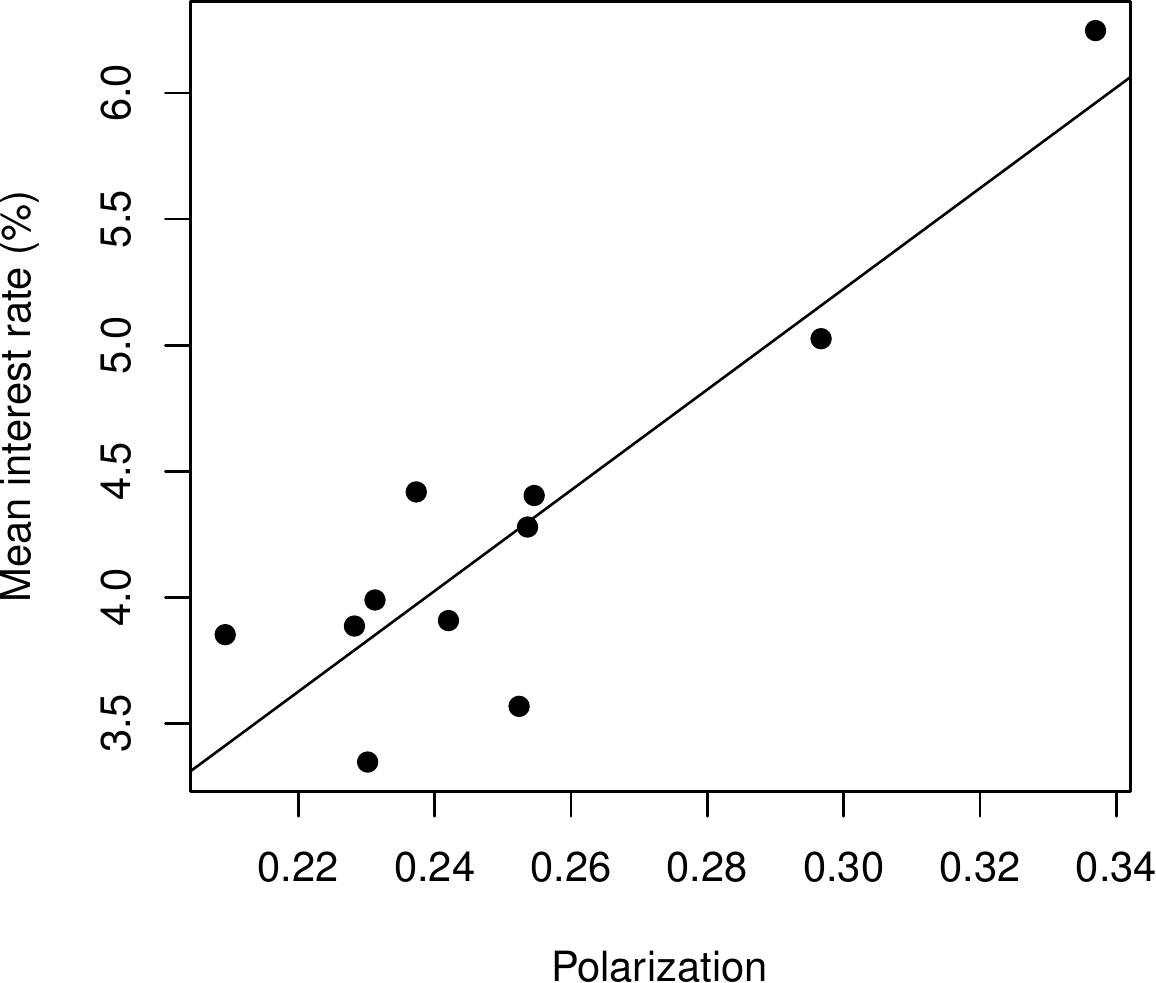}}
\vspace*{8pt}
\caption{Left: cross correlation between $Pol(t)$ and $Int(t+\Delta
   t)$, for $\Delta t \in [-9,9]$, where error bars show the 95\%
  confidence interval of the Pearson's correlation estimate. Right:
  scatter plot of $Pol(t)$  versus $Int(t+7)$, with linear regression result. 
  \label{fig:Debt-Pol-corr}}
\end{figure}

In our analysis, we focus on the 11 editions of Eurovision since 2002,
the year when the Euro was introduced as a physical currency. Figure
\ref{fig:Debt-Pol-TS} shows the time series of monthly mean interest
rate, and polarization in Eurovision. Both polarization and interest
rate jointly increase in 2010 and 2011, seemingly increasing
polarization before the interest rate. We analyze this joint movement
by calculating the cross correlation between both time series,
i.e. $\rho(Pol(t), Int(t+\Delta t))$, for values of $\Delta t$ between
$-9$ months and $+9$ months. The estimated values of $\rho(Pol(t),
Int(t+\Delta t))$ and their 95\% confidence intervals are shown in the
left panel of Figure \ref{fig:Debt-Pol-corr}, having a maximum value
of $0.894$ with p-value = $0.000205$ for $\Delta t = +7$. To deepen
more in the shape of this correlation, the right panel of Figure
\ref{fig:Debt-Pol-corr} shows the scatter plot of $Pol(t)$ and
$Int(t+7)$. The solid line shows the result of a linear regression of
the form $Int(Pol) = \alpha + \beta Pol$, where the estimates given by
least squares are $\alpha= -0.765$ and $\beta= 19.961$. This
regression has a residual error of $0.3775$ and an $R^2$ of $0.7995$,
explaining almost 80\% of the variance of the mean interest rate every
December.

While this analysis shows the existence of a correlation between both
time series, this does not imply the existence of a causal
relation. It seems unlikely that the results of a yearly cultural
event like Eurovision can influence the state of the economy of a
considerable part of Europe. Instead, the relation between these two
variables can be interpreted as support for the theory that both are
influenced by a third component, or that both are manifestations of
the same phenomenon. This way, the polarization in Eurovision would be
an early indicator for a social and cultural phenomenon, which is
followed by states of distrust in the economy of the involved
countries. To illustrate this theory, we show in Figure
\ref{fig:Debt-Pol-TS} the dates and amounts of loans from the
International Monetary Fund and the European Financial Stability
Facility, until January 2013. The increased polarization in the
contest is close in time to these events, being followed later by the
sovereign bond interest rate.

\subsection{Additional analysis}

The analysis explained above has two limitations: i) there is a free
parameter $\Delta t$ that needs to be accounted for, and ii) it is
only based on the polarization amount EU-15 countries, ignoring all
other countries and metrics of modularity discussed in Section
\ref{sec:modularity}. This means that we might fall into a \emph{Texas
  sharpshooter fallacy}, finding patterns when focusing on subsets of
random data. In the following, we deepen our analysis in order to
assess the robustness and limitations of our statistical results.

To control if the correlation between $Pol(t)$ and $Int(t+7)$ is due to
spurious fluctuations in the mean interest rate, we also calculated
the Pearson's correlation coefficient between the polarization and the
mean interest rate including all data in the following 7 months after
the contest. The result is a correlation coefficient of $0.692$ with
p-value = $0.01833$, showing that time averages also provide
significant correlations between both time series. In addition, the
correlation coefficients for all $\Delta t \geq 4$ were significant at
a 95\% confidence interval, highlighting the relation from past to
future that lies between polarization and mean interest rate.

The selection of polarization among the EU-15 countries needs to be
corrected, taking into account the familywise error rate of the whole
set of measurements. To control for this effect, we computed all the
correlations between the the mean interest rate after $\Delta t$
months, and the metrics of polarization, scores modularity $Q_s$, and
FoF modularity $Q_{fof}$, both for the EU-15 subnetwork and the whole
Eurovision dataset. Then we applied a Bonferroni correction of the
p-value of these correlations, computing a conservative estimate of
the probability of an incorrect rejection of the null hypothesis.

\begin{figure}[th]
\centerline{\includegraphics[width=0.85\textwidth]{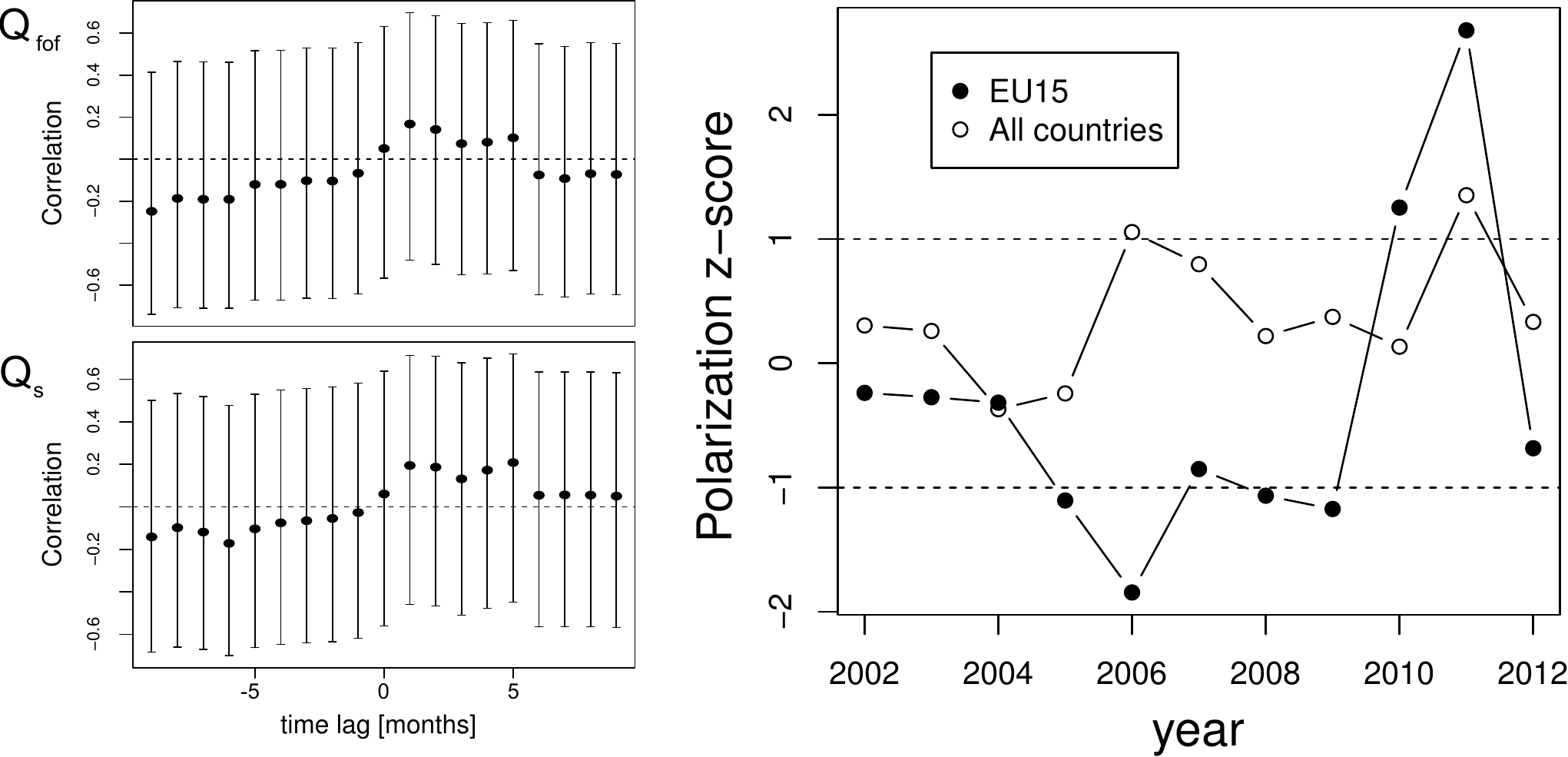}}
\vspace*{8pt}
\caption{ Left panel: time series of correlation coefficients between
  $Int(t+\Delta t)$ and EU-15 modularity for scores $Q_s$ and for FoF
  $Q_{fof}$. Right panel: time series of z-scores for the polarization
  of the EU-15 subnetwork and the whole Eurovision
  contest. \label{fig:Pol-Zscore}}
\end{figure}

Table \ref{tab:tab-corrs} shows the correlation estimate and the
original and corrected p-values for all the 6 metrics, taking $\Delta
t = 7$. After correction, the correlation between the polarization in
the EU-15 subnetwork and the mean interest rate is still
significant. The correlation between the polarization at the whole
Eurovision level can be initially accepted (p-value $=0.04231$), but
the correction reveals that the chance of being mistaken is much
higher (p-value $> 0.25$).  We can appreciate this effect by looking
into the time series of the z-score of both polarizations, $z(t) =
(Pol(t) - \mu_{Pol})/\sigma_{Pol}$. We calculated $\mu_{Pol}$ and
$\sigma_{Pol}$ for the time period between 1997 and 2012, comparing
the yearly value of both polarizations with their mean. The time
series of these z-scores is shown in the right panel of Figure
\ref{fig:Pol-Zscore}. While the z-score of the polarization for all
countries rarely goes beyond 1, the effect of increased polarization
in the EU-15 subnetwork is evident in 2010 and 2011.

\begin{table}[th]
\centering
\begin{tabular}{ l c c c c c }
  $var$ & set & $\rho(var,Int(t+7))$  & p-value &
  corrected p-value \\ \hline
  $Pol(t)$ & EU15  &  \textbf{0.894}   & $0.000205$  & \textbf{0.001230134} \\
  $Q_s(t)$ & EU15 &   $0.056$   & $0.8692$  & $1$ \\
  $Q_{fof}(t)$ & EU15  & $-0.092$       &  $0.7869$ & $1$\\
  $Pol(t)$ & All &  $0.619$  & $0.04231$  &  $0.253835928$ \\
  $Q_s(t)$ & All  &  $0.378$  & $0.251$ & $1$ \\
  $Q_{fof}(t)$ & All  & $-0.381$   &  $0.2477$  & $1$ \\
\end{tabular}
\caption{Pearson correlation coefficients between the different
  metrics discussed here and the lagged mean bond interest rate
  $Int(t+7)$, 95\% confidence intervals, and p-values with Bonferroni correction. \label{tab:tab-corrs}}
\end{table}

All the other variables did not show any significant correlation,
leaving the polarization of EU-15 as the best metric to study the
relation between culture in Eurovision and the economy in the EU. For
other values of $\Delta t$, the results are similar as explained
above, i.e. the correlation with EU-15 polarization is only
significant when $\Delta t \geq 5$, and the other metrics keep showing
low significance. The left panel of Figure \ref{fig:Pol-Zscore} shows
the cross correlation between the mean interest rate and the EU-15
modularity for FoF, and for scores. We did not find any value for
$\Delta t$ that supports the assumption that there is some relation
between the financial crisis and a possible division of the EU-15
countries in Eurovision.

\section{Discussion}

We have studied the cultural relations among European countries
through the behavioral biases present in the Eurovision song contest.
To do so, we gathered a dataset of the historical contest outcomes,
which aggregates the votes of large amounts of viewers who
simultaneously vote by phone calls and messages.  Our approach is
centered around the statistical analysis of this large-scale dataset,
producing metrics that compose a macroscope of the cultural cohesion
in Europe at large.

The first metric defined here is the Friend-or-Foe coefficient, a
metric that reveals asymmetric positive and negative relations between
European countries. We validated how this metric represents cultural
affinity by comparing its historical values with previous data on
European cultures. This result is consistent with previous research
\cite{Ginsburgh2008}, where the influence of culture was taken into
account to predict individual votes.  In a more general setup, our
findings suggest that there is a relation between cultural distance
and voting biases that prevails through time, showing the relation
between known cultural relations and the FoF.  Furthermore, we
designed a model of affinity in Eurovision contests that, when
compared to null models, shows the influence of cultural affinity in
the FoF.

Using the FoF on empirical data, we found a community structure with
higher modularity than the equivalent using votes alone. The
modularity of this FoF network keeps approximately constant through
time, while the modularity of the raw scores network increases every
year. This suggests that Eurovision participants adapt within the
contest rules, in contrast to their cultural biases, which seem to be
define constant subcommunities according to their cultures.

Over this network data, we designed the metric of polarization, which
detects changes in the voting patterns among participating
countries. Applying this metric to the votes between countries in the
EU-15, we find a significant change in 2010 and 2011, the most
turbulent years with respect to debt and austerity measures in the EU.
Our empirical analysis of the correlation between polarization and EU
debt indicators supports the hypothesis that both are related to the
political climate in the EU.  This suggests, in turn, that political
decisions can influence the perception of culture across countries,
and economic decisions of the involved states can change the way
societies relate to each other in the EU.

While we find that there is a positive lag between polarization and
sovereign bond interest rates, one can argue against the usage of
Eurovision polarization as a predictor for interest rates. High states
of polarization coocur with bailouts, which can be seen as the
precursors of changes in the interest rates. Nevertheless, our
polarization metric provides a way to quantify how \emph{society}
reacts to political decisions and the crisis in general, in a similar
manner as sovereign bond interest rates measure how \emph{the market}
reacts to the same phenomenon.

Additional datasets are available to extend our
results. \texttt{Wikipedia} also contains the results of the
semifinals of the contest, which can be used to refine the
Friend-or-Foe coefficient as an aggregation of all the available
data. Furthermore, our structural analysis can be combined with
statistics of online behavior, such as measures of search queries,
website visits, \texttt{Twitter} posts, or amounts of views and
comments for the \texttt{Youtube} videos of participants. These
extended datasets, in turn, could be used to create testable
predictors for the outcome of Eurovision contests.

A clear limitation of our analysis is the country level restriction
given the data provided by the Eurovision song contest. Cultures do
not need to map to countries, as ethnic minorities or pan-state
cultures are neglected in this analysis. This limitation is present in
the current state-of-the-art studies \cite{Ailon2008}, waiting for
sources of cultural data at different levels of aggregation.  Our
estimation of cultural affinity through the Friend-or-Foe coefficient
depends on the way these two relate to each other, which we explored
through model simulations. 

The modeling and analysis approach presented here can be applied to
other contests, for example in artist popularity competitions
\cite{Ciulla2012}, or beauty contests. The Friend-or-Foe coefficient
can be adapted to other contest schemes, looking for alternative
support on the way cultural affinity creates voting biases. Finally,
modularity and polarization can be used to create an``European mood''
metric, which relates to the economical and political decisions of the
European Union, and its member states. Applying this metric to future
contests, we can investigate how political decisions influence the
state of cultural cohesion between the inhabitants of European
countries.

\section*{Acknowledgments}
We would like to thank Uwe Serd\"ult, Andreas Flache, and Frank
Schweitzer for providing useful comments and discussions.

\end{document}